\documentclass[conference,letterpaper]{IEEEtran}
\usepackage[top=0.73in,bottom=1.05in,left=0.637in,right=0.64in]{geometry}
\IEEEoverridecommandlockouts

\usepackage{enumerate}
\usepackage[utf8]{inputenc} 
\usepackage[T1]{fontenc}
\usepackage{url}
\usepackage{ifthen}
\usepackage{tcolorbox}
\usepackage[noadjust]{cite}
\usepackage[cmex10]{amsmath} 
\usepackage{amssymb,amsfonts,amscd,amsbsy}
\usepackage{graphicx}
\usepackage{stmaryrd} % for \llbracket and \rrbracket
\usepackage[mathscr]{eucal}
\usepackage{mathtools}
\usepackage{bbm}
\usepackage{xcolor}
\usepackage{xfrac}
\usepackage{algorithm}
\usepackage{algpseudocode}
\usepackage{enumerate}
\usepackage{enumitem}
\usepackage{multirow}
\newtheorem{theorem}{Theorem}
\numberwithin{theorem}{section}
\newtheorem{corollary}[theorem]{Corollary}
\newtheorem{lemma}[theorem]{Lemma}
\newtheorem{definition}[theorem]{Definition}
\usepackage{etoolbox}
\newcommand{\overbar}[1]{\mkern 1.5mu\overline{\mkern-1.5mu#1\mkern-1.5mu}\mkern 1.5mu}

\makeatletter
\renewcommand*{\thesection}{\arabic{section}}
\renewcommand*{\thesubsection}{\thesection.\arabic{subsection}}
\renewcommand*{\p@subsection}{}
\renewcommand*{\thesubsubsection}{\thesubsection.\arabic{subsubsection}}
\renewcommand*{\p@subsubsection}{}
\makeatother

\usepackage{titlesec}
\titleformat{\subsection}[runin]
  {\normalfont\normalsize\bfseries}{\thesubsection.}{3pt}{}[.]
\titleformat{\subsubsection}[runin]
  {\normalfont\normalsize\itshape}{\thesubsubsection.}{3pt}{}[\mbox{~}]

\newcommand{\suppress}[1]{}

\newcommand{\Fb}{\mathbbmss{F}} % field
\newcommand{\Zb}{\mathbbmss{Z}} % integers
\newcommand{\Cs}{\mathscr{C}}   % code
\newcommand{\wt}{{\rm w}_{\!H}}   % weight
\newcommand{\RM}{\mathrm{RM}}   % Reed-Muller code
\newcommand{\tth}{\text{th}}    % such as i^th --> $i^\tth$
\newcommand{\p}{\pmb}           % bold font
\newcommand{\dmin}{d_{\min}}    % min distance

\newcommand{\dimenof}{{\rm dim}}  % dimension
\newcommand{\spanof}{{\rm span}}  % span
\newcommand{\Wc}{\mathscr{W}}
\newcommand{\abelian}{\mathscr{C}_{\rm A}} % abelian code with (m,W)
\newcommand{\bid}{{\rm BiD}}
\newcommand{\bidd}{{\rm BiD}^\perp}

\newcommand{\Rm}{\mathscr{R}}

\newcommand{\punc}{{\rm punc}}

\title{On Decoding First- and Second-Order BiD Codes}
\author{Devansh Jain and Lakshmi Prasad Natarajan%
\thanks{The authors are with the Department of Electrical Engineering, Indian Institute of Technology Hyderabad, Sangareddy 502248, India (email: \{ee22btech11018,  lakshminatarajan\}@iith.ac.in).}%
\thanks{This work was supported by the Qualcomm 6G University Research India Program.}% <-this % stops a space
}

\begin{document}

\maketitle

\begin{abstract}
BiD codes, which are a new family of algebraic codes of length $3^m$, achieve the erasure channel capacity under bit-MAP decoding and offer asymptotically larger minimum distance than Reed-Muller (RM) codes. 
In this paper we propose fast maximum-likelihood (ML) and max-log-MAP decoders for first-order BiD codes. 
For second-order codes, we identify their minimum-weight parity checks and ascertain a code property known as `projection' in the RM coding literature. 
We use these results to design a belief propagation decoder that performs within $1$~dB of ML decoder for block lengths $81$ and $243$.
% 
% we use the minimum-weight parity checks and the projection property to design a belief propagation decoder.
% and a belief propagation decoder for second-order codes
% We use the idea of code projections and the minimum-weight codewords to design a belief propagation (BP) decoder for second-order BiD codes. For lengths $81$ and $243$, the block error rate of the BP decoder is within $1.00$~dB of the maximum-likelihood decoder.
\end{abstract}

% \begin{IEEEkeywords}
% decoding, minimum weight codewords, product codes, Reed-Muller codes, weight distribution.
% \end{IEEEkeywords}

% % % % % 
% include this in arXiv
% \clearpage
% \setcounter{tocdepth}{2} % do not show subsubsections
% \tableofcontents
% \clearpage

% % % % %

\section{Introduction}

Reed-Muller (RM) codes~\cite{Muller_IRE_54,Reed_IRE_54,ASY_IT_20}, 
% which are one of the earliest constructed and extensively studied families of error correcting codes, 
have recently been proved to achieve the capacity of binary-input memoryless symmetric (BMS) channels~\cite{AbS_FOCS_2023,ReP_IT_2024}. 
Compared to Polar codes, RM codes offer an explicit construction with rich structure, larger minimum distances and lower block error rates (BLER) under maximum-likelihood (ML) decoding~\cite{MHU_TCOM_2014}.
% There are multiple related ways to construct RM codes~\cite{ASY_IT_20,Ber_Cybernetics_67} that illuminate the rich structure and symmetry of these codes: such as the $(\p{u}\,|\,\p{u}+\p{v})$ Plotkin construction or from polynomial evaluations.
% the rows of the matrix $\p{A}_2^{\otimes m}$ (where $\p{A}_2=[1~0;\,1~1]$ is the Ar{\i}kan's $2 \times 2$ kernel matrix). 
Apart from the fact that good low-complexity decoders are not known, RM codes face a major practical limitation---unlike Polar codes~\cite{KSU_IT_10,BGLB_TCOM_20}, their block lengths are limited to powers of $2$.
% % The rich structure of RM codes are illuminated by the multiple methods by which these codes can be constructed~\cite{ASY_IT_20,Ber_Cybernetics_67}
% \emph{(i)}~using evaluations of polynomials over $\Fb_2$; 
% \emph{(ii)}~the $(\p{u} \, | \, \p{u} + \p{v})$ or Plotkin construction;
% \emph{(ii)}~from the rows of the matrix $\p{A}_2^{\otimes m}$, where $\p{A}_2= [1~0;\,1~1]$ is Ar{\i}kan's $2 \times 2$ kernel matrix and $\otimes$ is the Kronecker product; and 
% \emph{(iii)}~as an ideal in the group algebra of the group $(\Zb_2^m,+)$.
% These constructions show the rich structure and symmetry of RM codes. However, this richness is achieved at the cost of a major practical limitation---the block lengths of RM codes are limited to powers of $2$.

Berman-intersection-Dual Berman (BiD) codes~\cite{DNNK_ITW_25} are a family of algebraic codes constructed for lengths \mbox{$N=3^m$} that achieve vanishingly small bit error rates in the binary erasure channel at all rates less than capacity. 
For any chosen code rate $R \in (0,1)$, there exist BiD codes whose minimum distance grows at least as fast as $N^{0.543}$ as $N \to \infty$. This is asymptotically larger than the distance of RM codes (which grows as $\sqrt{N}$). 
% Simulations~\cite{DNNK_ITW_25} show that the BLER (under ML decoding) of some BiD codes are comparable to or better than RM and CRC-aided Polar codes. 
Simulations of some BiD codes~\cite{DNNK_ITW_25} exhibit BLERs better than or comparable to RM and CRC-aided Polar codes under ML decoding.

BiD codes might prove useful in constructing a good family of algebraic codes with flexible lengths. For instance, BiD codes could be used with an RM construction technique (such as the Plotkin construction) to obtain codes of lengths $2^{m_1}3^{m_2}$ for all \mbox{$m_1,m_2 \geq 0$}, and hopefully, with minimum distance or BLER comparable to RM codes. 
While this direction appears promising, we presently investigate more proximate questions---analyzing the dual codes of and designing low-complexity decoders for some BiD codes. 

% Contributions
In this paper, we first review the construction and properties of BiD codes from~\cite{DNNK_ITW_25} (Section~\ref{sec:bid_codes}). We then identify all the minimum-weight parity checks of second-order BiD codes (Section~\ref{sec:min_wt_parity_checks}). 
Next we show that a well-known property called \emph{projection}, which is used by several RM decoding algorithms~\cite{YeA_IT_20,LHP_ISIT_20,DuS_IT_06,SiP_PPI_92,Sak_ITW_05}, is enjoyed by BiD codes as well (Section~\ref{sec:projection}). We use these new results together with ideas from~\cite{YeA_IT_20,LHP_ISIT_20,SHP_ISIT_18,BeS_IT_86,SNK_ISIT24} to design efficient decoders for first- and second-order BiD codes (Section~\ref{sec:decoding}).
Simulation results show that the proposed belief propagation (BP) decoder for second-order BiD codes of lengths $81$ and $243$ perform within $1$~dB of the ML decoder, and the block error rate (BLER) is similar to that of CRC-aided Polar codes under successive cancellation list decoding with list size $8$ (Section~\ref{sec:simulations}).
% Several known decoding algorithms for RM codes~\cite{YeA_IT_20,DuS_IT_06,SiP_PPI_92,Sak_ITW_05,LHP_ISIT_20} rely on the fact that the sum of two carefully selected sub-vectors of any codeword of $\RM(m,r)$ lies in $\RM(m-1,r-1)$. 
% We show that this property, which is known as \emph{projection}, applies to the family of BiD codes too (Section~\ref{sec:projection}). 

\emph{Notation:} 
% The symbol $\otimes$ denotes the Kronecker product. For any positive integer $\ell$ 
% let $[\ell] = \{1,\dots,\ell\}$, and 
Let $\Zb_{\ell} \triangleq \{0,1,\dots,\ell-1\}$ be the ring of integers modulo $\ell$.
% Let $\Zb_{\ell}=\lset \ell \rset$ be the ring where addition and multiplication are performed modulo $\ell$.
% The binary field is $\Fb_2=\{0,1\}$. 
% The empty set is $\emptyset$.
% For sets $A, B$, $A \setminus B \triangleq \left\{a \in A~:~ a \notin B \right\}$, and $\bar{A}$ is the complement of $A$. 
For a set $A$, $\bar{A}$ is its complement.
Capital bold letters denote matrices,  small bold letters denote row vectors, $(\cdot)^T$ is the transpose operator, and $\wt$ denotes the Hamming weight.
% We use $\left[\,\p{a}_1; \, \p{a}_2; \, \cdots ; \, \p{a}_\ell\,\right]$ to denote the matrix with rows $\p{a}_1,\dots,\p{a}_{\ell}$.
% All vectors are row vectors, unless otherwise stated. 
% The notation $(.)^T$ denotes the transpose operator.
% The notation $\p 0$ denotes a zero-vector of appropriate size. 
% We denote the identity matrix of size $n$ by ${\p I}_n$. 
% For two vectors ${\p a},{\p b}$, their concatenation is denoted by $({\p a}|{\p b}).$  
The dimension, minimum distance and dual code of a linear code (subspace) $\mathcal{C}$ are $\dimenof(\mathcal{C})$, $\dmin(\mathcal{C})$ and $\mathcal{C}^\perp$, respectively.
% The dimension of a linear code (subspace) $\mathcal{C}$ is denoted by $\dimenof(\mathcal{C})$, its minimum distance by $\dmin(\mathcal{C})$ and its dual code by $\mathcal{C}^\perp$. 
% The Hamming weight of a vector ${\p a}$ is denoted by $\wt({\p a})$. 
% The support of a vector $\p{a}$ is $\supp(\p{a})$.
% The minimum distance of a code ${\Cs}$ is denoted by $d_{\min}({\Cs}).$ 
% We use $\spanof$ to denote the span of a collection of vectors, and 
% $\rowspaceof$ to denote the span of the rows of a matrix.
% The binomial coefficient $\binom{n}{k}$ is assumed to be $0$ if $k>n$ or if $k<0$. 
% Finally, $\indicator$ denotes the indicator function.

\section{BiD Codes: Construction \& Properties} \label{sec:bid_codes}

% We will now review two techniques to construct BiD codes from~\cite{DNNK_ITW_25}, one algebraic and the other recursive. These constructions will be used to analyze projection properties of the BiD code and the minimum weight codewords of its dual code.

% \subsection{Algebraic Construction} 

% We first review the framework used in the construction of abelian codes as ideals in the group algebra of the group $(\Zb_3^m,+)$, and then define BiD codes within this framework~\cite{DNNK_ITW_25}.

% We first review the algebraic framework used in the construction of BiD codes and then recall the definition of these codes from~\cite{DNNK_ITW_25}.

We review the algebraic framework, definition and properties of BiD codes from~\cite{DNNK_ITW_25}.

\subsection{Algebraic Framework}

Consider the polynomial quotient ring 
% \begin{equation*}
$\Rm = \Fb_2[X_1,\dots,X_m] / \langle X_1^3-1,\dots,X_m^3-1 \rangle$    
% \end{equation*}
in $m$ indeterminates $X_1,\dots,X_m$. As an algebra over $\Fb_2$ the dimension of $\Rm$ is $3^m$ and the monomials 
% \begin{equation*}
$\left\{ X_1^{i_1} \cdots X_m^{i_m}:i_1,\dots,i_m \in \Zb_3 \right\}$
% \end{equation*}
% where \mbox{$\Zb_3=\{0,1,2\}$}, 
form an $\Fb_2$-basis of $\Rm$. 
For any $\p{i} = (i_1,\dots,i_m) \in \Zb_3^m$ we will use $X^{\p{i}}$ to denote the monomial $X_1^{i_1} \cdots X_m^{i_m}$. 
Every polynomial \mbox{$f \in \Rm$} can be expressed as \mbox{$f = \sum_{\p{i} \in \Zb_3^m} f_{\p{i}} X^{\p{i}}$} where ${f}_{\p{i}} \in \Fb_2$ is the coefficient of the monomial $X^{\p{i}}$.
We associate with $f$ the vector \mbox{$\p{f} = (f_{\p{i}} : \p{i} \in \Zb_3^m)$} of length $3^m$. The entries of $\p{f}$ are arranged in the increasing order of $\sum_{\ell=1}^{m} i_{\ell} 3^{\ell-1}$.
Thus, $f$, $f_{\p{i}}$ and $\p{f}$ denote a polynomial, its $\p{i}^\tth$ coefficient, and the vector composed of all its coefficients, respectively.

% \subsubsection*{Fourier transform.}
The $m$-dimensional discrete Fourier transform (DFT)~\cite{Nic_JCSS_71,RaS_IT_92} of the binary vector $\p{f}$ is a length-$3^m$ vector $\p{\hat{f}}$ over the field $\Fb_4=\{0,1,\alpha,\alpha^2\}$. Like $\p{f}$, the entries of \mbox{$\p{\hat{f}} = (\hat{f}_{\p{j}} \,: \,\p{j} \in \Zb_3^m)$} are indexed by the vectors in $\Zb_3^m$. These are related to the components of $\p{f}$ as $\hat{f}_{\p{j}} = \sum_{\p{i} \in \Zb_3^m} f_{\p{i}} \alpha^{\p{i} \cdot \p{j}}$ where $\p{i} \cdot \p{j} = \sum_{\ell=1}^{m} i_{\ell} j_{\ell}$ is the dot product of $\p{i}$ and $\p{j}$ over $\Zb_3$. 
Using the fact \mbox{$\alpha^3=1$} it is easy to see that $\hat{f}_{\p{j}}$ is the evaluation of the polynomial $f$ at the point $(\alpha^{j_1},\dots,\alpha^{j_m}) \in \left(\Fb_4^* \right)^m$ where $\Fb_4^*=\{1,\alpha,\alpha^2\}$.
The evaluations of $f$ at all the points in $(\Fb_4^*)^m$ forms its spectral description $\p{\hat{f}}$. 
The binary vector $\p{f}$ can be obtained from $\p{\hat{f}}$ via \mbox{$f_{\p{i}} = \sum_{\p{j} \in \Zb_3^m} \hat{f}_{\p{j}} \alpha^{-\p{i} \cdot \p{j}}$}. 
The DFT $f \to {\p{\hat{f}}}$ embeds the $\Fb_2$-algebra $\Rm$ into $\Fb_4^{3^m} = \oplus_{\p{j} \in \Zb_3^m} \Fb_4$. In particular, we have the \emph{convolution property}: if $f,g,h \in \Rm$, then $fg=h$ in $\Rm$ if and only if $\hat{f}_{\p{j}}\hat{g}_{\p{j}} =  \hat{h}_{\p{j}}$ for every $\p{j} \in \Zb_3^m$.

\subsection{Codes from $\Rm$}

% BiD codes are a subclass of a family of abelian codes $\abelian$ constructed from the ring $\Rm$. 
For any integer \mbox{$m \geq 1$} and \emph{frequency weight set} $\Wc \subseteq \{0,1,\dots,m\}$, we define an abelian code
\begin{equation} \label{eq:abelian_code}
\abelian(m,\Wc) \triangleq \left\{ \,f \in \Rm \, : \, \hat{f}_{\p{j}} = 0 \, \forall \, \p{j} \text{ with } \wt(\p{j}) \notin \Wc \right\}.
\end{equation}
That is, $f$ lies in the code if $(\alpha^{j_1},\dots,\alpha^{j_m})$ is a root of $f$ for every $\p{j}$ with $\wt(\p{j}) \notin \Wc$. 
From the convolution property we observe that if $f \in \abelian$, $g \in \Rm$ and \mbox{$fg=h$}, then for all $\wt(\p{j}) \notin \Wc$ we obtain $\hat{h}_{\p{j}}=\hat{f}_{\p{j}}\hat{g}_{\p{j}}=0$, and thus, \mbox{$h \in \abelian$}.
Hence $\abelian$ is an ideal in $\Rm$.
With some abuse of notation, we use $\abelian$ to denote not only this ideal in $\Rm$ but also the corresponding linear block code $\{\p{f} \,:\, f \in \abelian \} \subseteq \Fb_2^{3^m}$.

The dimension of $\abelian(m,\Wc)$ is $\sum_{w \in \Wc} 2^w \binom{m}{w}$ which is the number of vectors in $\Zb_3^m$ with Hamming weight in $\Wc$.
Its dual code is $\abelian(m,\overbar{\Wc})$, where $\overbar{\Wc}=\{0,1,\dots,m\} \setminus \Wc$. 
For any $\Wc,\Wc'$,
\emph{(i)}~$\abelian(m,\Wc \cap \Wc') = \abelian(m,\Wc) \cap \abelian(m,\Wc')$; and 
\emph{(ii)}~$\abelian(m,\Wc) \subset \abelian(m,\Wc')$ if and only if $\Wc \subset \Wc'$. 
Further, $\abelian(m,\Wc)$ is the direct-sum of its subcodes $\abelian(m,\{w\}), w \in \Wc$, that is, $\abelian(m,\Wc)= \bigoplus_{w \in \Wc} \abelian(m,\{w\})$. 

% \subsection{On the Automorphism Group}

% Section~IV-D of~\cite{NaK_IT_23} identifies certain automorphisms of $\abelian(m,\Wc)$. %, and hence, of BiD codes and their duals too. 
% We repeat this analysis here for the sake of completeness. 
% Later, we will use these automorphisms to identify all the minimum-weight parity checks of $\bid(m,2,2)$.
% 
% An automorphism is a permutation of the coordinates of a vector that maps every codeword to a codeword. 
The coordinates of our codewords are indexed by the monomials $X^{\p{i}}$, and hence, we view code automorphisms as permutations on the monomial basis \mbox{$\{X^{\p{i}} : \p{i} \in \Zb_3^m\}$}.
For completeness we have included a proof of the following result~\cite[Section~IV-D]{NaK_IT_23} in Appendix~\ref{app:lem:automorphisms}.

\begin{lemma} \label{lem:automorphisms}
For any \mbox{$m \geq 1$} and \mbox{$\Wc \subseteq \{0,\dots,m\}$} the following permutations are automorphisms of $\abelian(m,\Wc)$
\begin{itemize}
\item[\emph{(i)}] for any $\p{k} \in \Zb_3^m$, the map $X^{\p{i}} \to X^{\p{i} + \p{k}}$ for all $\p{i}$; % where the addition $\p{i} + \p{k}$ is over $\Zb_3$;
\item[\emph{(ii)}] for any $\ell \in \{1,\dots,m\}$, replacing $X_{\ell}$ with $X_{\ell}^2$; and  
\item[\emph{(iii)}] for any permutation $\gamma$ on $\{1,\dots,m\}$, replacing $X_{\ell}$ with $X_{\gamma(\ell)}$ for all $\ell \in \{1,\dots,m\}$.
\end{itemize}
\end{lemma}
% \begin{IEEEproof}
% For completeness wwe have included a proof in Appendix~\ref{app:lem:automorphisms}.
% \end{IEEEproof}
We do not know if the above generate the entire automorphism group. 
The automorphisms identified in Lemma~\ref{lem:automorphisms} satisfy the conditions laid out in~\cite[Theorem~19]{KCP_ISIT16}; hence, asymptotically long abelian codes $\abelian$ attain vanishingly small bit error rates in the binary erasure channel at all rates less than capacity~\cite{NaK_IT_23}.

\subsection{BiD Codes}

The family of BiD codes are those abelian codes whose frequency weight sets are contiguous integers. BiD codes are parameterized by integers $m,r_1,r_2$ with $0 \leq r_1 \leq r_2 \leq m$. We refer to $r_1$ and $r_2$ as the \emph{order parameters}. 
\begin{definition}
The BiD code $\bid(m,r_1,r_2)$ is the abelian code $\abelian(m,\Wc)$ with $\Wc=\{r_1,r_1+1,\dots,r_2\}$.
\end{definition}

The BiD code with $\Wc=\{r_1,\dots,m\}$ is the $(r_1-1)^\tth$-\emph{order Berman code}~\cite[Corollary~32]{NaK_IT_23}. The Berman code was first designed in~\cite[proof of Theorem~2.1]{Ber_Cybernetics_II_67} for lengths $p^m$ for any odd prime $p$ (not just \mbox{$p=3$} as considered in this paper). 
The code with \mbox{$\Wc=\{0,\dots,r_2\} = \{0,\dots,m\} \setminus \{r_2+1,\dots,m\}$} is the dual of the $r_2^\tth$-order Berman code. % The minimum distances of the Berman code $\bid(m,r_1,m)$ and the dual Berman code $\bid(m,0,r_2)$, are $2^{r_1}$ and $3^{m-r_2}$, respectively~\cite{Ber_Cybernetics_II_67,BlN_IT_01,NaK_IT_23}. 
For \mbox{$r_1 \leq r_2$}, the intersection of these two codes is precisely the Berman-intersection-Dual Berman code $\bid(m,r_1,r_2)$, and its $\dmin$ is lower bounded~\cite{DNNK_ITW_25} by % its minimum distance is lower bounded by~\cite{DNNK_ITW_25} 
% \begin{equation*}
$\max\{4^{r_1}3^{m-r_1-r_2},\, 3^{m-r_2}2^{r_1 + r_2 - m}\}$.
% \end{equation*}
This lower bound is tight for \mbox{$r_1=0$} (any $r_2$), \mbox{$r_2=m$} (any $r_1$), \mbox{$r_1=r_2=1$} and \mbox{$r_1=r_2=m-1$}, see~\cite{Ber_Cybernetics_II_67,BlN_IT_01,NaK_IT_23,DNNK_ITW_25}.
Drawing parallels between RM and BiD codes, we refer to $\bid(m,1,1)$ and $\bid(m,2,2)$ as the first- and second-order BiD codes, respectively; see Table~\ref{table:first_second_order_RM_BiD} for comparison with RM codes. 

\begin{table}[t]
\renewcommand{\arraystretch}{1.25}
\caption{Comparison of the dimension $K$ and the distance $\dmin$ of\\ first- and second-order RM and BiD codes.}
\label{table:first_second_order_RM_BiD}
\centering
\begin{tabular}{l|c|c}
\hline 
\hline 
Code & $K$ & $\dmin$ \\
\hline 
\hline 
$\RM(m,1)$ & $\approx \log_2 N$ & $0.5 N$ \\
\hline
$\bid(m,1,1)$ &  $\approx 1.26 \log_2 N$ & $0.44 N$ \\
\hline 
\hline 
$\RM(m,2)$ & $\approx 0.5 \log_2^2 N$ & $0.25 N$ \\
\hline 
$\bid(m,2,2)$ & $\approx 0.79 \log_2^2 N$ & $\geq 0.19 N$ \\
\hline
\hline 
\end{tabular}
\vspace{-4mm}
\end{table}

% \subsection{Recursive Construction}

% Applying the direct-sum decomposition property of $\abelian(m,\Wc)$ to BiD codes, we have $\bid(m,r_1,r_2) = \bigoplus_{w=r_1}^{r_2} \bid(m,w,w)$. Hence, the generator matrix of $\bid(m,r_1,r_2)$ can be obtained by vertically stacking the generator matrices of $\bid(m,w,w)$. Now, Lemma~2.3 of~\cite{DNNK_ITW_25} gives a recursive construction of the generator matrix of $\bid(m,w,w)$ in terms of the generator matrices of BiD codes of length $3^{m-1}$.

\section{Minimum-Weight Parity Checks of $\bid(m,2,2)$} \label{sec:min_wt_parity_checks}

% The parity check constraints on a code are precisely the non-zero codewords from its dual code.
% In this section we identify the minimum distance and all the minimum-weight codewords of $\bidd(m,2,2)$, $m \geq 4$; later, we will use these as check nodes in the BP decoding of $\bid(m,2,2)$.

We now identify all the minimum-weight codewords of $\bidd(m,2,2)$, $m \geq 3$. They serve as minimum-weight parity checks for second-order BiD codes.

\subsection{Main Idea}
Suppose $\abelian(m,\Wc)$ contains a non-zero codeword $f$ of Hamming weight\footnote{Unlike the other sections of this paper, where $w$ represents an element of $\Wc$, here we use $w$ to represent a Hamming weight.} $w$.
% If the code $\abelian(m,\Wc)$ contains a codeword of Hamming weight\footnote{Unlike the other sections where $w$ denotes an element of $\Wc$, in this section we use $w$ to denote the Hamming weight of a vector.} $w \neq 0$ 
There exist $w$ distinct vectors \mbox{$\p{a}_1, \dots, \p{a}_w \in \Zb_3^m$} such that \mbox{$f = \sum_{i=1}^{w} X^{\p{a}_i}$}. 
% If \mbox{$f \in \Rm$} has Hamming weight\footnote{Unlike the other sections where $w$ denotes an element of $\Wc$, in this section we use $w$ to denote the Hamming weight of a vector.} $w$ then there exist $w$ distinct vectors $\p{a}_1, \dots, \p{a}_w \in \Zb_3^m$ such that $f = \sum_{i=1}^{w} X^{\p{a}_i}$. If $f$ is a codeword of an abelian code $\abelian(m,\Wc)$, 
% the automorphisms from Lemma~\ref{lem:automorphisms} apply; in particular, 
Using the automorphism of type~\emph{(i)} from Lemma~\ref{lem:automorphisms} on $f$ with $\p{k} = -\p{a}_w$ yields a codeword in $\abelian(m,\Wc)$ such that one of its non-zero terms is \mbox{$X^{\p{0}}=1$}.
Hence, without loss of generality, we will assume that \mbox{$\p{a}_w=\p{0}$}, i.e., \mbox{$f = 1 + \sum_{i=1}^{w-1} X^{\p{a}_i}$} where $\p{a}_1,\dots,\p{a}_{w-1}$ are distinct and non-zero. Further, we will assume that $\p{a}_1,\dots\p{a}_{w-1}$ are in the lexicographic order.

Now consider the specific case of $\bidd(m,2,2)$ which is the code $\abelian(m,\Wc)$ with $\Wc=\{0,1,\dots,m\} \setminus \{2\}$. If $f = 1 + \sum_{i=1}^{w-1} X^{\p{a}_i}$ lies in this code, from~\eqref{eq:abelian_code}, we have 
\begin{equation} \label{eq:constraint_on_a_i}
\textstyle 
\sum_{i=1}^{w-1} \alpha^{\p{a}_i \cdot \p{j}}=1 \text{ for all } \p{j} \in \Zb_3^m \text{ with } \wt(\p{j})=2.
\end{equation}
Clearly, codewords with \mbox{$w=1$} do not exist since the LHS above evaluates to $0$.
% \mbox{$w = 1$} is not a possibility since the LHS above evaluates to $0$. Henceforth we will assume \mbox{$w \geq 2$}.
% for all $\p{j} \in \Zb_3^m$ with $\wt(\p{j})=2$.
Next, we rewrite~\eqref{eq:constraint_on_a_i} as constraints on the columns of a \mbox{$(w-1) \times m$} matrix $\p{M}$ over $\Zb_3$. 
Let $\p{M}$ be the matrix with rows $\p{a}_1,\dots,\p{a}_{w-1}$ (in that order) and let $V = \left\{\p{j} \in \Zb_3^m : \wt(\p{j}) = 2 \right\}$. 
The \mbox{$w-1$} exponents of $\alpha$ in the LHS of~\eqref{eq:constraint_on_a_i} are precisely the components of the row vector $\p{j}\p{M}^T$ computed over $\Zb_3$. 
Thus \mbox{$\p{j} \p{M}^T \in S$} for every \mbox{$\p{j} \in V$} where
\begin{equation} \label{eq:S_definition}
\textstyle S = \left\{ (u_1, \dots, u_{w-1}) \in \Zb^{w-1}_3 : \sum^{w-1}_{i=1}\alpha^{u_i} = 1 \right\}.
\end{equation}
We summarize this observation as 
\begin{lemma} \label{lem:min_wt_reformulation}
% $\bidd(m,2,2)$ contains a weight-$w$ codeword, $w \geq 2$, if and only if there exists an $\p{M} \in \Zb_3^{(w-1) \times m}$ with distinct non-zero rows such that $\p{j}\p{M}^T \in S$ for all $\p{j} \in V$. 
% Further, there is one-to-one correspondence between the set of $\p{M}$ satisfying this condition and codewords of weight 
% For \mbox{$w \geq 2$}, the weight-$w$ codewords in $\bidd(m,2,2)$ that contain $X^{\p{0}}$ as one of their terms are in a one-to-one correspondence with $\{ \p{M} \in \Zb_3^{(w-1) \times m} : \p{j} \p{M}^T \in S \,\, \forall \p{j} \in V \}$.
Let $\mathscr{M}$ be the collection of all \mbox{$\p{M} \in \Zb_3^{(w-1) \times m}$} with non-zero and distinct rows which are arranged in lexicographic order such that $\p{j}\p{M}^T \in S$ for all $\p{j} \in V$. The weight-$w$ codewords, $w \geq 2$, of $\bidd(m,2,2)$ that contain $X^{\p{0}}$ as one of their terms are in a one-to-one correspondence with $\mathscr{M}$.
\end{lemma}

Lemma~\ref{lem:min_wt_reformulation} is a collection of constraints on the columns of $\p{M}$. Using $\p{c}_{\ell}^T$ to denote the $\ell^\tth$ column of $\p{M}$, and considering all possible $\p{j} \in V$, we see that 
\begin{equation} \label{eq:min_wt_reformulation_columns}
a_1 \p{c}_{\ell_1} + a_2 \p{c}_{\ell_2} \in S     
\end{equation}
for every choice of $a_1,a_2 \in \Zb_3 \setminus \{0\}$ and $1 \leq \ell_1 < \ell_2 \leq m$.

% For a codeword of weight $w$ to exist in $\bidd(m,2,2)$, there must exist $w$ distinct vectors $\p{a}_1, \dots, \p{a}_w \in \Zb_3^m$ such that:
% \begin{equation*}
%     0 = \sum_{i=1}^w \alpha^{\p{a}_i \cdot \p{j}}
% \end{equation*}
% for all $\p{j} \in \Zb_3^m$ with $\wt(\p{j}) = 2$. By the first automorphism defined in Lemma~\ref{lem:automorphisms}, we can set $\p{a}_w = \p{0}$ and $\tilde{\p{a}}_i = \p{a}_i - \p{a}_w$. The condition simplifies to:
% \begin{equation*}
%     1 = \sum_{i=1}^{w-1} \alpha^{\p{a}_i \cdot \p{j}}
% \end{equation*}
% Consider the $m \times (w-1)$ matrix $M = \left[\,\p{a}_1; \, \p{a}_2; \, \cdots ; \, \p{a}_{w-1}\,\right]$. Let $V = \{\p{v} \in \Zb_3^m : \wt(\p{v}) = 2\}$. For a codeword of weight $w$ to exist, there must be a matrix $M$ with distinct non-zero rows such that for every $\p{v} \in V$, the multiplication $M\p{v} \in S$, where:
% \begin{equation*}
%     S = \{ (u_1, \dots, u_{w-1}) \in \Zb^{w-1}_3 : \sum^{w-1}_{i=1}\alpha^{u_i} = 1 \}
% \end{equation*}

% \subsection{Bounding the Distance of $\bidd(m,2,2)$}

Lemmas~\ref{lem:automorphisms} and~\ref{lem:min_wt_reformulation} imply that if $\mathscr{M}$ is empty for some choice of $w$ then no codewords of weight $w$ exist in $\bidd(m,2,2)$. 
We use this idea and the constraints~\eqref{eq:min_wt_reformulation_columns} to show that for $m \geq 3$ there are no codewords of weights $1,\dots,4$ in $\bidd(m,2,2)$. This gives us

\begin{lemma} \label{lem:bidd_lower_bound_5}
The minimum distance of $\bidd(m,2,2)$, $m \geq 3$, is at least $5$.
\end{lemma}

\emph{Proof Idea.} 
We prove the non-existence of $\p{M}$ for $w=2,3,4$ using~\eqref{eq:min_wt_reformulation_columns} and the fact that $S \subset \Zb_3^{w-1}$ is closed under negation. For $w=4$, we also use the fact that $S$ is a union of three linearly independent lines in $\Zb_3^3$. See Appendix~\ref{app:lem:bidd_lower_bound_5} for proof.
% We use the structure of $S$ within the vector space $\Zb_3^{w-1}$; in particular, the fact that $S$ is closed under negation. For $w=4$, we also use the fact that $S$ is a union of three linearly independent lines in $\Zb_3^3$. We use these observations to prove the non-existence of corresponding $\p{M}$; see Appendix~\ref{app:lem:bidd_lower_bound_5}. 
% \hfill \IEEEQED

% \begin{IEEEproof}
% Please see Appendix~\ref{app:lem:bidd_lower_bound_5}.
% \end{IEEEproof}

% An immediate consequence of Lemma~\ref{lem:bidd_lower_bound_5} is

\begin{corollary} \label{cor:bidd_322_dmin}
The minimum distance of $\bidd(3,2,2)$ is $5$.
\end{corollary}
\begin{IEEEproof}
It is enough to show that $\bidd(3,2,2)$ contains a codeword of weight $5$. Direct verification via~\eqref{eq:constraint_on_a_i} shows that $1 + X_1X_2X_3 + X_1X_2^2X_3^2 + X_1^2X_2X_3^2 + X_1^2X_2^2X_3$ is one such codeword.
\end{IEEEproof}

We verified that the application of all permutations generated by automorphisms of types~\emph{(i)} and~\emph{(ii)} of Lemma~\ref{lem:automorphisms} on $1 + X_1X_2X_3 + X_1X_2^2X_3^2 + X_1^2X_2X_3^2 + X_1^2X_2^2X_3$ yields all the $54$ minimum-weight codewords of $\bidd(3,2,2)$.

\subsection{Minimum-Weight Parity Checks} \label{sec:sub:min_wt_parity_checks}

Unless otherwise stated we will assume that \mbox{$m \geq 4$} in this subsection.
The lower bound in Lemma~\ref{lem:bidd_lower_bound_5} happens to be loose for these values of $m$; in these cases the minimum distance is $6$. 
We will arrive at this result---along with the complete characterization of all the minimum-weight codewords---using Lemma~\ref{lem:min_wt_reformulation} and an inexpensive computer search. 
The main idea is to work with a simple graph (undirected, no loops) $G = (V_G, E_G)$ where the vertex set $V_G = \Zb_3^{w-1}$ is the collection of vectors from which we draw columns for $\p{M}$ and the edge set $E_G$ is the reflection of the constraint~\eqref{eq:min_wt_reformulation_columns}. 
For $\p{c},\p{c'} \in V_G$ we have $\{\p{c},\p{c'}\} \in E_G$ if and only if $a\p{c} + a'\p{c'} \in S$ for all $a,a' \in \{1,2\}$.

\subsubsection*{Minimum Distance of $\bidd(m,2,2)$.}
We first show that there are no weight-$5$ codewords.
For $w=5$, the set $S$ consists of all permutations of the vectors $(k, k, 1, 2)$ for $k \in \{0, 1, 2\}$. Since $\p{0} \notin S$ and $\p{c}_{\ell_1} - \p{c}_{\ell_2} \in S$ for all $\ell_1 \neq \ell_2$, the matrix $\p{M}$ must consist of distinct columns. 
Following the definition of $E_G$, we observe that $\{\p{c}_1,\dots,\p{c}_m\}$ will form a clique of size $m$ in $G$.
A computer search for maximal cliques in $G$ gives a largest clique of size $3$. 
Since $m>3$ no such matrix $\p{M}$ exists, and thus, there are no codewords of weight $5$ in $\bidd(m,2,2)$. Hence, we have $\dmin \geq 6$ for this code. 
The bound $\dmin \leq 6$ follows from the direct observation that the polynomial 
\begin{equation} \label{eq:weight_6_polynomial}
f = (X_1 + X_1^2)(1 + X_2\cdots X_m + X_2^2\cdots X_m^2)    
\end{equation}
of weight $6$ satisfies $\hat{f}_{\p{j}}=0$ for all $\p{j}$ with $\wt(\p{j})=2$.

% We use graph theory to search for valid $M$. Let $G = (V_G, E_G)$ be a graph where the vertex set $V_G = \Zb_3^m \setminus \{\p{0}\}$ represents all possible non-zero columns. An edge exists between two distinct columns $c_i$ and $c_j$ if and only if all their linear combinations $a c_i + b c_j$ for $a, b \in \{1, 2\}$ belong to the set $S$. Here, a valid matrix $M$ corresponds to a clique in $G$. 
% A computer search for maximal cliques in $G$ gives a largest clique of size $3$. 
% Since $m>3$ no such matrix $\p{M}$ exists and thus there are no codewords of weight $5$ in $\bidd(m,2,2)$.

% Since we can't construct a valid matrix with more than three columns, a codeword of weight $w=5$ does not exist for $\bidd(m,2,2)$ for $m > 3$. However the minimum distance for $\bidd(3,2,2)$ is 5.

% Computational search for maximal cliques in $G$ gives a largest clique of size 3. Since we can't construct a valid matrix with more than three columns, a codeword of weight $w=5$ does not exist for $\bidd(m,2,2)$ for $m > 3$. However the minimum distance for $\bidd(3,2,2)$ is 5.

\subsubsection*{Minimum-Weight Codewords of $\bidd(m,2,2)$.}

For $w=6$, $S$ consists of all permutations of the vectors $(0,a,a,b,b)$ for $a,b \in \{0,1,2\}$. 
Since $S$ contains $\p{0}$ the columns of $\p{M}$ need not be distinct. However, if $\p{c} \in \Zb_3^5$ is such that two or more columns of $\p{M}$ are equal to $\p{c}^T$, then applying~\eqref{eq:min_wt_reformulation_columns} on two such columns we have $\p{c}, 2\p{c} \in S$.

In order to relate an $\p{M} \in \mathscr{M}$ to the graph $G$, we pick the unique columns of $\p{M}$ and identify them with the corresponding vertices in $V_G=\Zb_3^5$. These nodes (which might number less than $m$ because of column repetitions in $\p{M}$) must then form a clique in $G$. 
If the size of this clique is indeed less than $m$, it means that at least one of the elements of this clique appears as two or more columns in $\p{M}$. Thus, at least one of the elements of such a clique must be `repeatable', i.e.,  must belong to $S$. 
This process can map multiple matrices $\p{M}$ to the same clique.

We performed a computer search for cliques in $G$ that correspond to $\p{M}$ with distinct and non-zero rows. The largest such clique is of size $3$, which is less than $m$ (since $m \geq 4$ for us). 
Thus, any clique of interest to us must contain at least one repeatable node. 
A computer search for such maximal cliques in $G$ yields the following two cliques
\begin{align} \label{eq:cliques_C1_C2}
\begin{split}
C_1 &= \left\{ 
(0,0,1,1,1), (1,2,0,1,2), (2,1,0,2,1)
\right\}, \text{ and } \\
C_2 &= \left\{ 
(0,0,2,2,2), (1,2,0,1,2), (2,1,0,2,1)
\right\}.
\end{split}
\end{align}
% \begin{equation} \label{eq:cliques_C1_C2}
% C_1 = \left\{ 
% \begin{bmatrix}
% 0 \\ 0 \\ 1 \\ 1 \\ 1 
% \end{bmatrix},
% \begin{bmatrix}
% 1 \\ 2 \\ 0 \\ 1 \\ 2 
% \end{bmatrix},
% \begin{bmatrix}
% 2 \\ 1 \\ 0 \\ 2 \\ 1 
% \end{bmatrix}
% \right\}
% ,
% C_2 = \left\{ 
% \begin{bmatrix}
% 0 \\ 0 \\ 2 \\ 2 \\ 2 
% \end{bmatrix},
% \begin{bmatrix}
% 1 \\ 2 \\ 0 \\ 1 \\ 2 
% \end{bmatrix},
% \begin{bmatrix}
% 2 \\ 1 \\ 0 \\ 2 \\ 1 
% \end{bmatrix}
% \right\}.
% \end{equation}
In both these cliques the last two vectors are repeatable.
For constructing $\p{M}$ from either $C_1$ or $C_2$, the first vector must be used as a column exactly once (to ensure that $\p{M}$ has non-zero rows). Each of the other two vectors in the clique can be repeated any number of times (including zero times) and the columns can be permuted to arrive at a valid $5 \times m$ matrix $\p{M} \in \mathscr{M}$. 
For $C_1$, we have $2^{m-1}$ duplication patterns of the last two vectors followed by $m$ distinct placements of the first vector as columns of $\p{M}$. Since we are only interested in lexicographic ordering of rows, we observe that the roles played by the second and third vectors in $C_1$ are interchangeable. This yields us $m2^{m-1}/2$ codewords from $C_1$. We get the same number from $C_2$ as well, yielding a total of $m2^{m-1}$ minimum-weight codewords with $X^{\p{0}}$ as one of the terms.

% After a computational search for maximal cliques and filtering for matrices with distinct non-zero rows (while removing row permutations), the largest identified clique is of size 3.  To create a matrix with $m \ge 4$ columns, we must have duplicate columns; a column $c_i$ is repeatable only if $c_i \in S$. Exhaustive construction of all possible submatrices containing at least one repeatable column gives the following four sub-matrices:
% \begin{equation*}
% \left[ \!\!
% \begin{array}{cc}
% 0 & 1\\
% 0 & 2\\
% 1 & 0\\
% 1 & 1\\
% 1 & 2
% \end{array}
% \!\! \right],
% \left[ \!\!
% \begin{array}{cc}
% 0 & 1\\
% 0 & 2\\
% 2 & 0\\
% 2 & 1\\
% 2 & 2
% \end{array}
% \!\! \right],
% \left[ \!\!
% \begin{array}{ccc}
% 0 & 1 & 2\\
% 0 & 2 & 1\\
% 1 & 0 & 0\\
% 1 & 1 & 2\\
% 1 & 2 & 1
% \end{array}
% \!\! \right],
% \left[ \!\!
% \begin{array}{ccc}
% 0 & 1 & 2\\
% 0 & 2 & 1\\
% 2 & 0 & 0\\
% 2 & 1 & 2\\
% 2 & 2 & 1
% \end{array}
% \!\! \right]
% \end{equation*}
% From the $5 \times2$ matrices, we can create $2m$ distinct $5\times m$ matrices by duplicating columns and considering all column permutations. For the $5\times3$ matrix, swapping repeatable columns gives the same matrix. After duplicating columns and considering all column permutations, we can create $m(2^{m-1} - 2)$ matrices. Together, this gives $m2^{m-1}$ codewords with support containing the zero vector $(\p{a}_w =\p{0})$.

Any other minimum-weight codeword can be obtained from these $m2^{m-1}$ codewords by applying a permutation of type~\emph{(i)} from Lemma~\ref{lem:automorphisms}, i.e., by translating the support set of the codewords by $\p{k} \in \Zb_3^m$. 
For any codeword $f = 1 + \sum_{i=1}^{5} X^{\p{a}_i}$ with $\p{0}$ in its support, there are $6$ choices of $\p{k}$ (viz., $0,2\p{a}_1,\dots,2\p{a}_5$) such that even the translated codeword has $\p{0}$ in its support. 
Thus, allowing all $\p{k} \in \Zb_3^m$, this process covers each minimum-weight codeword exactly $6$ times. 
Hence, we have proved
\begin{theorem} \label{thm:bidd_dmin}
The minimum distance of $\bidd(m,2,2)$, $m \geq 4$, is $6$ and the number of weight-$6$ codewords is $m2^{m-2}3^{m-1}$.
\end{theorem}

% Using the first automorphism defined in Lemma~\ref{lem:automorphisms}, we shift these supports to generate $3^m$ codewords from each support. However if $\p{k}$ is twice of any row, we obtain a codeword already identified in the zero support set. Therefore we will obtain a total of $2^{m-2}3^{m-1}m$ distinct codewords of weight $w=6$.

% \subsubsection*{Polynomial Form of Weight-$6$ Codewords.}

In Appendix~\ref{app:polynomial_form_wt_6_codewords} we use a counting argument to show that all the minimum-weight codewords of $\bidd(m,2,2)$, $m \geq 4$, can be obtained by applying the automorphisms from Lemma~\ref{lem:automorphisms} on the polynomial~\eqref{eq:weight_6_polynomial}.
% 
% We have identified a polynomial of weight $w = 6$ for $\bidd(m,2,2)$:
% \begin{equation*}
% f = (X_1 + X_1^2)(1 + X_2\dots X_m + X_2^2\dots X_m^2)
% \end{equation*}

The weight-$6$ codewords of $\bidd(m,2,2)$ do not span this code. In fact, the linear code defined by using these weight-$6$ vectors as parity checks is $\bid(m,2,2) \oplus \bid(m,0,0)$, where $\bid(m,0,0)$ is the repetition code of length $3^m$ (the proof, which is available in Appendix~\ref{app:span_of_min_wt_codewords}, uses the structure of ideals in semisimple rings~\cite{Ber_Cybernetics_II_67,MacWilliams_Bell_70,RaS_IT_92}). 
Hence, a BP decoder built using these parity checks decodes to the super-code $\bid(m,2,2) \oplus \bid(m,0,0)$ instead of $\bid(m,2,2)$.

% The set of vectors that satisfy all the constraints imposed by these minimum-weight parity checks is not $\bid(m,2,2)$ but its super-code $\bid(m,2,2) \oplus \bid(m,0,0)$, where $\bid(m,0,0)$ is the repetition code of length $3^m$. 
% We prove this result in Appendix~\ref{app:span_of_min_wt_codewords} by showing that the span of these parity checks is an ideal and by using the structure of ideals in semisimple rings~\cite{Ber_Cybernetics_II_67,MacWilliams_Bell_70,RaS_IT_92}.

% this span is an ideal in $\Rm$ and by using the structure theorem of ideals~\cite{Ber_Cybernetics_II_67,MacWilliams_Bell_70,RaS_IT_92} in the semisimple ring $\Rm$.

% The span of these minimum-weight codewords is a subcode of $\bidd(m,2,2)$ (with one dimension less). We prove this result in Appendix~\ref{app:span_of_min_wt_codewords} by showing that this span is an ideal in $\Rm$ and by using the structure theorem of ideals~\cite{Ber_Cybernetics_II_67,MacWilliams_Bell_70,RaS_IT_92} in the semisimple ring $\Rm$. 
% This result implies that the constraints imposed by the minimum-weight parity checks are satisfied by all codewords in the super-code $\bid(m,2,2) \oplus \bid(m,0,0)$, where $\bid(m,0,0)$ is the repetition code of length $3^m$.

\subsection{On the Dual Distance of Other BiD Codes}

In~\cite[Section~3]{DNNK_ITW_25} an efficient recursive algorithm was presented to compute upper and lower bounds on the minimum distance of an arbitrary BiD code $\bid(m,r_1,r_2)$. 
The key ideas used in this technique are: 
\emph{(i)}~the recursive formulation~\cite[Lemma~2.3]{DNNK_ITW_25} of the generator matrix of $\abelian(m,\{w\})$ in terms of $\abelian(m-1,\{w\})$ and $\abelian(m-1,\{w-1\})$; and
\emph{(ii)}~partitioning a codeword $\p{\rho}$ into three sub-vectors $\p{\rho}_0,\p{\rho}_1,\p{\rho}_2$ and relating $\wt(\p{\rho})$ to the weights of $\p{\rho}_i$, $i=0,1,2$.
Fortuitously, this technique from~\cite{DNNK_ITW_25} works just as it is for any abelian code $\abelian(m,\Wc)$. 
% since it does not rely on the peculiarities of BiD codes beyond the fact that~\cite[Lemma~2.3]{DNNK_ITW_25} applies to them.
% 
In Appendix~\ref{app:dual_distance_recursion} we revisit the recursive technique of~\cite{DNNK_ITW_25} in the context of applying the idea to any abelian code $\abelian(m,\Wc)$, and then we present the numerical results for computing bounds on the minimum distance of all dual BiD codes $\bidd(m,r_1,r_2)$, $0 \leq r_1 \leq r_2 \leq m$. 
We use Corollary~\ref{cor:bidd_322_dmin} and Theorem~\ref{thm:bidd_dmin} as termination conditions in this numerical recursion. 
Thus, these results have consequences beyond the duals of second-order BiD codes.
% these bounds on $\dmin$ of any $\bidd(m,r_1,r_2)$ are consequences of Corollary~\ref{cor:bidd_322_dmin} and Theorem~\ref{thm:bidd_dmin}.

% Since the proof of~\cite[Theorem~3.1]{DNNK_ITW_25} applies almost verbatim to this more general setting (with only straightforward modifications to account for the larger family of codes), we keep our presentation as concise as possible.

\section{Projection Property of BiD Codes} \label{sec:projection}

% We will now use the algebraic construction of BiD codes to analyze their projection properties.
% Please note that these results can also be proved using the recursive structure~\cite[Lemma~2.3]{DNNK_ITW_25} of their generator matrices.

% \subsection{Puncturing BiD Codes}

Before introducing projections for BiD codes we analyze what puncturing does to the Fourier coefficients.
Consider \mbox{$\p{f} = (f_{\p{i}}\,:\,\p{i} \in \Zb_3^m)$}, where $f_{\p{i}} = \sum_{\p{j} \in \Zb_3^m} \hat{f}_{\p{j}} \alpha^{-\p{i} \cdot \p{j}}$, and $\hat{f}_{\p{j}}$ are the spectral coefficients.
Suppose we want to puncture $\p{f}$ to obtain a sub-vector $\p{a}$ of length $3^{m-\ell}$. We do so by fixing the last $\ell$ coordinates $i_{m-\ell+1},\dots,i_{m}$ of the index $\p{i}=(i_1,\dots,i_{m})$ to some constants, say, $u_1,\dots,u_{\ell} \in \Zb_3$, respectively.
That is, % define the punctured vector $\p{a}=(a_{\p{i'}}\,:\,\p{i'} \in \Zb_3^{m-\ell})$ as 
\begin{equation*}
a_{\p{i'}} = f_{(\p{i'},\p{u})} = f_{(i_1',\dots,i_{m-\ell}',u_1,\dots,u_{\ell})} \text{ for all } \p{i'} \in \Zb_3^{m-\ell}.
\end{equation*}
Now, $\p{a}=(a_{\p{i'}}:\p{i'} \in \Zb_3^{m-\ell})$ can be described using its \mbox{$(m-\ell)$}-dimensional DFT $\hat{a}_{\p{j'}}$, $\p{j'} \in \Zb_3^{m-\ell}$. 
For $\p{j} \in \Zb_3^m$, let us denote the first $m-\ell$ components of $\p{j}$ as $\p{j'} = (j_1,\dots,j_{m-\ell})$ and the remaining $\ell$ components as $\p{j''}$. Then we have ${a}_{\p{i'}} = f_{(\p{i'},\p{u})} = \sum_{\p{j} \in \Zb_3^m} \hat{f}_{\p{j}} \alpha^{-\left(\p{i'} \cdot \p{j'} + \p{u} \cdot \p{j''} \right)}$, that is,
\begin{equation*}
% \textstyle {a}_{\p{i'}} &= f_{(\p{i'},\p{u})} = \sum_{\p{j} \in \Zb_3^m} \hat{f}_{\p{j}} \alpha^{-\left(\p{i'} \cdot \p{j'} + \p{u} \cdot \p{j''} \right)} \\
\textstyle
{a}_{\p{i'}} = \sum_{\p{j'} \in \Zb_3^{m-\ell}} \left( \sum_{\p{j''} \in \Zb_3^{\ell}} \hat{f}_{\p{j}} \alpha^{-\p{u} \cdot \p{j''}} \right) \alpha^{-\p{i'} \cdot \p{j'}}
\end{equation*}
% Denoting $(j_1,\dots,j_{m-1})$ as $\p{j'}$, we have 
% \begin{equation*}
% \textstyle
% {a}_{\p{i'}} = \sum_{\p{j'} \in \Zb_3^{m-1}} \left( \sum_{j_m \in \Zb_3} \hat{f}_{\p{j}} \alpha^{-uj_m} \right) \alpha^{-\p{i'} \cdot \p{j'}},
% \end{equation*}
which is precisely the inverse DFT map for $\p{a}$. Thus, we have 
\begin{equation} \label{eq:puncurting_DFT}
\textstyle
\hat{a}_{\p{j'}} = \sum_{\p{j''} \in \Zb_3^{\ell}} \hat{f}_{\p{j}} \alpha^{-\p{u} \cdot \p{j''}}.  
\end{equation}
% 
% \begin{lemma} \label{lem:puncturing}
% With $\p{f},\p{a}$ as defined above and $w \geq \ell$, if $\p{f} \in \bid(m,w,w)$ then $\p{a} \in \bid(m-\ell,w-\ell,w)$.
% \end{lemma}
% \begin{IEEEproof}
% The first $m-\ell$ entries of $\p{j} \in \Zb_3^m$ constitute $\p{j'}$. 
Suppose $\p{f} \in \bid(m,w,w)$.
If $\wt(\p{j'}) \notin \{w-\ell,\dots,w\}$ then irrespective of the value of $\p{j''}$ we have $\wt(\p{j}) \neq w$. That is, if $\wt(\p{j'}) \notin \{w-\ell,\dots,w\}$, we have $\hat{f}_{\p{j}} = 0$ for all $\p{j''} \in \Zb_3^{\ell}$, and hence, from~\eqref{eq:puncurting_DFT}, we have $\hat{a}_{\p{j'}} = 0$. Thus, we have $\p{a} \in \bid(m-\ell,w-\ell,w)$ 
% \textcolor{red}{$\bid(m-\ell,w-\ell,w)$} 
if $\p{f} \in \bid(m,w,w)$.
% \end{IEEEproof}

% \subsection{Projection for BiD Codes}

% So far, we punctured the length-$3^m$ vector $\p{f}$ to length-$3^{m-\ell}$ sub-vector $\p{a}$ by setting the last $\ell$ entries of the index $\p{i}$ to a constant $\p{u}$. 
Now, for any \mbox{$\p{v} \in \Zb_3^{\ell} \setminus \{\p{u}\}$}, consider the sub-vector $\p{b}$ of $\p{f}$ obtained by setting $(i_{m-\ell+1},\dots,i_m)=\p{v}$. 
% Clearly, the sub-vectors $\p{a}$ and $\p{b}$ have no coordinates of $\p{f}$ in common.
Define the \emph{projection} $\p{p} = \p{a} + \p{b}$, i.e., $p_{\p{i'}} = a_{\p{i'}} + b_{\p{i'}}$ for all $\p{i'} \in \Zb_3^{m-\ell}$.
% The sum $\p{p}$ of the sub-vectors $\p{a}$ and $\p{b}$ is a \emph{projection}. 
From~\eqref{eq:puncurting_DFT}, the Fourier coefficients of $\p{p}$ are
\begin{equation*}
\textstyle 
\hat{p}_{\p{j'}} = \sum_{\p{j''} \in \Zb_3^{m-\ell}} \hat{f}_{\p{j}} \left( \alpha^{-\p{u} \cdot \p{j''}} + \alpha^{-\p{v} \cdot \p{j''}} \right) ~\forall~\p{j'} \in \Zb_3^{m-\ell}.
\end{equation*}
% This observation immediately gives us

\begin{lemma} \label{lem:projection}
With $\p{f},\p{p}$ as defined above and $w \geq \ell$, if $\p{f} \in \bid(m,w,w)$ then $\p{p} \in \bid(m-\ell,w-\ell,w-1)$.
\end{lemma}
\begin{IEEEproof}
% Following the same argument as in the proof of Lemma~\ref{lem:puncturing}, we see that 
It is clear that $\hat{p}_{\p{j'}} = 0$ if $\wt(\p{j'}) \notin \{w-\ell,\dots,w\}$.
Further, if $\wt(\p{j'})=w$, then $\hat{f}_{\p{j}} \neq 0$ only if $\p{j''}=\p{0}$. However, in this case $\hat{f}_{\p{j}} ( \alpha^{-\p{u} \cdot \p{j''}} + \alpha^{-\p{v} \cdot \p{j''}} ) = \hat{f}_{\p{j}}(1+1) = 0$. 
Thus, $\hat{p}_{\p{j'}} = 0$ when $\wt(\p{j'}) \notin \{w-\ell,\dots,w-1\}$.
\end{IEEEproof}

We can generalize Lemma~\ref{lem:projection} by considering projections where any set of $\ell$ coordinates of the index $\p{i}$ are fixed (not necessarily the last $\ell$ coordinates).
% The automorphisms of type~\emph{(iii)} in Lemma~\ref{lem:automorphisms} allow us to cover this generalization.
% 
% in two different ways: 
% \emph{(i)}~puncture by fixing the value of any coordinate of $\p{i}$ (not necessarily $i_m$); and
% \emph{(ii)}~puncture by fixing more than one coordinate of $\p{i}$.
% The last set of automorphisms identified in Lemma~\ref{lem:automorphisms} allow us to cover the generalization~\emph{(i)}, while the ideas used in the proofs of Lemmas~\ref{lem:puncturing} and~\ref{lem:projection} continue to work for~\emph{(ii)}. 
% In this paper, we use projections to decode $\bid(m,2,2)$, and hence, we now state only those generalizations that are relevant to this task. 
% Let us introduce some notation towards this generalization. 
Let $S = \{s_1,\dots,s_{\ell}\} \subset \{1,\dots,m\}$ denote the coordinates of the index $\p{i}$ that will be fixed to constant values, and let $\p{u} \in \Zb_3^{\ell}$ denote these constants. 
We will use $\punc(\p{f},S,\p{u})$ to denote the sub-vector of $\p{f}$ consisting of those indices $\p{i}$ that satisfy $i_{s_1}=u_1,\dots,i_{s_{\ell}}=u_{\ell}$.

\begin{theorem} \label{thm:projection}
Let $S \subset \{1,\dots,m\}$ with $|S|=\ell$, $1 \leq \ell \leq w$, $\p{u},\p{v} \in \Zb_3^{\ell}$ and $\p{u} \neq \p{v}$. For any $\p{f} \in \bid(m,w,w)$ we have 
\begin{equation*}
\punc(\p{f},S,\p{u}) + \punc(\p{f},S,\p{v}) \in \bid(m-\ell,w-\ell,w-1).
\end{equation*}
\end{theorem}
\begin{IEEEproof}
We can use a code automorphism (of type~\emph{(iii)} from Lemma~\ref{lem:automorphisms}) to map $X_{s_1},\dots,X_{s_{\ell}}$ to $X_{m-\ell+1},\dots,X_{m}$, respectively. This maps the index components $i_{s_1},\dots,i_{s_{\ell}}$ to $i_{m-\ell+1},\dots,i_{m}$, respectively. The required result then follows from Lemma~\ref{lem:projection}.
% Please see Appendix~\ref{app:thm:projection}.
% We can assume \mbox{$S=\{m-\ell+1,\dots,m\}$}, because otherwise we can use an appropriate code automorphism (of type~\emph{(iii)} from Lemma~\ref{lem:automorphisms}) to map $X_{s_1},\dots,X_{s_{\ell}}$ to $X_{m-\ell+1},\dots,X_{m}$, respectively. This maps the index components $i_{s_1},\dots,i_{s_{\ell}}$ to $i_{m-\ell+1},\dots,i_{m}$, respectively.
% With $S=\{m-\ell+1,\dots,m\}$, this theorem is same as Lemma~\ref{lem:projection}.
\end{IEEEproof}

We will use projections with \mbox{$\ell=1,2$} in the BP decoding of $\bid(m,2,2)$ codes. 
Since these projections lie in the codes $\bid(m-1,1,1)$ and $\bid(m-2,0,1)$, respectively, we need soft-in soft-out decoders for these two families of codes.

\section{Decoding First- and Second-Order Codes} \label{sec:decoding}

\subsection{Decoding $\bid(m,1,1)$ and $\bid(m,0,1)$}

Fast implementations of ML decoders and soft-in soft-out max-log-MAP decoders exist for these two codes (see Appendix~\ref{app:decoding_first_order}). 
These efficient implementations are possible because the structure of $\bid(m,1,1)$ and $\bid(m,0,1)$ are similar to a family of codes known as \emph{recursive subproduct codes}~\cite{SNK_ISIT24}. In Appendix~\ref{app:decoding_first_order} we show how the efficient ML decoder and the fast max-log-MAP decoder of first-order recursive subproduct codes from~\cite{SNK_ISIT24} can be readily adapted to these two classes of $\bid$ codes. 
The idea behind these decoders is similar to the fast Hadamard transform based decoding technique for $\RM(m,1)$ codes~\cite{BeS_IT_86}.
% 
% \begin{remark} \label{rem:maxlogMAP_complexity}
Based on the analysis from~\cite{SNK_ISIT24}, the fast ML decoder and fast max-log-MAP decoders of $\bid(m,1,1)$ and $\bid(m,0,1)$ have complexity order $N^{2/\log_2 3} = N^{1.26}$, as against the complexity order $N^{2.26}$ of the naive ML decoder.   
% \end{remark}

\subsection{Belief Propagation Decoding of $\bid(m,2,2)$} \label{sec:sub:BP_decoding}

We use projections (with $\ell=1,2$) and the minimum-weight codewords of $\bidd(m,2,2)$ for BP decoding of second-order BiD codes. 
Our technique is similar to the decoders proposed in~\cite{LHP_ISIT_20} and~\cite{SHP_ISIT_18} for RM codes.
% We use Fig.~2(b) and Algorithm~1 of~\cite{LHP_ISIT_20} as reference points to describe our BP decoder for $\bid(m,2,2)$.

The factor graph of our decoder contains \mbox{$N=3^m$} variable nodes and $m 2^{m-2} 3^{m-1}$ degree-$6$ check nodes (corresponding to all the minimum-weight parity checks).
In addition, and similar to~\cite[Fig.~2(b)]{LHP_ISIT_20}, the factor graph contains hidden variable nodes (representing projected codewords) and generalized check nodes (representing soft-in soft-out decoders for the projections).
We use all possible projections with \mbox{$\ell=1$} from Theorem~\ref{thm:projection} which give us $3m$ generalized check nodes for decoding projections in $\bid(m-1,1,1)$.
Finally, we include all projections with \mbox{$\ell=2$} where \mbox{$\p{u}=(u_1,u_2)$} and \mbox{$\p{v}=(v_1,v_2)$} satisfy \mbox{$u_1 \neq v_1$} and \mbox{$u_2 \neq v_2$}; we do not consider cases where one of the components of $\p{u}$ and $\p{v}$ are identical since the corresponding \mbox{$\ell=2$} projection is already included as a sub-vector of some projection with $\ell=1$.
The number of generalized check nodes corresponding to \mbox{$\bid(m-2,0,1)$} in our factor graph is $18 \binom{m}{2}$.

We describe the finer details of the BP decoder in Appendix~\ref{app:bp_decoding}. The per-iteration complexity order of this decoder is $N^{\log6 / \log 3} \log N=N^{1.63} \log N$, which is the order of the number of minimum-weight parity checks of $\bid(m,2,2)$.

\section{Simulation Results} \label{sec:simulations}

We present simulation results for first- and second-order BiD codes in the binary-input AWGN channel. 
% We compare these codes with their closest RM codes and CRC-aided Polar codes.

\subsection{First-Order Codes}

\begin{figure}[t]
    \centering
    \includegraphics[width=\linewidth]{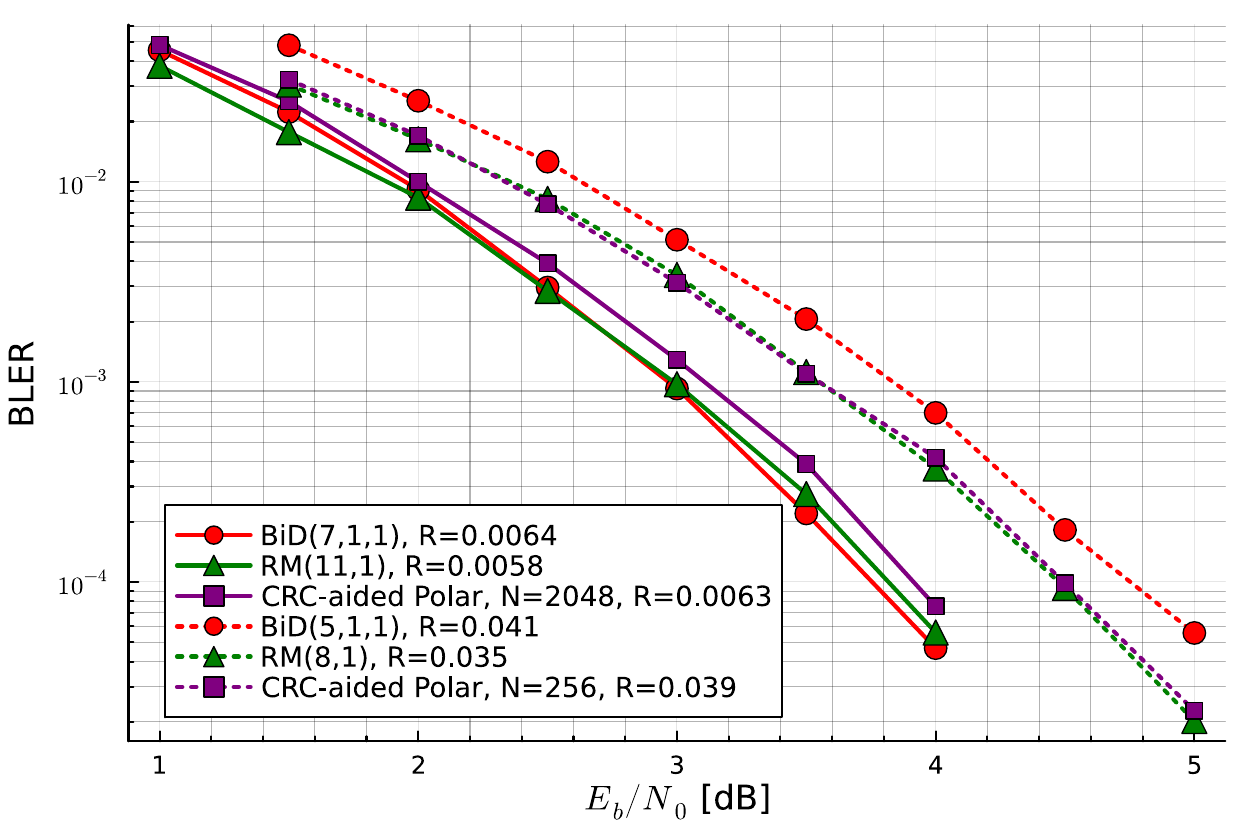}
    \caption{Comparison of first-order BiD and RM codes (ML decoding).}
    \label{fig:first_order}
\end{figure}

We compare first-order BiD codes with first-order RM codes and CRC-aided Polar codes (all under ML decoding) under two regimes of block length, $N \approx 250$ and \mbox{$N \approx 2000$}, which are represented as dashed and bold lines, respectively in Fig.~\ref{fig:first_order}. 
The BiD codes are decoded using the fast ML decoder adapted from the decoder of first-order recursive subproduct codes~\cite{SNK_ISIT24} (see Appendix~\ref{app:decoding_first_order}). The RM codes are decoded using the fast Hadamard transform based technique~\cite{BeS_IT_86}.
We use the successive cancellation ordered search (SCOS) decoder~\cite{YuC_TCOM_24,YuC_github} to perform an essentially-ML decoding for CRC-aided Polar codes (with $8$-bit CRC).

The parameters of the codes in Fig.~\ref{fig:first_order} with $N \approx 250$ are as follows:  
\emph{(i)}~$\bid(5,1,1)$ is a $[243,10,108]$ code;
\emph{(ii)}~$\RM(8,1)$ is a $[256,9,128]$ code; and
\emph{(iii)} the CRC-aided Polar code is a $[256,10]$ code.
For $N \approx 2000$, we have 
\emph{(i)}~$\bid(7,1,1)$ (a $[2187,14,972]$ code);
\emph{(ii)}~$\RM(11,1)$ ($[2048,12,1024]$); and
\emph{(iii)} the $[2048,13]$ CRC-aided Polar code. 
The rates of these codes are available in the legend box of Fig.~\ref{fig:first_order}.
We observe that the BLER of BiD codes, which have a slightly higher rate, are similar to those of RM and CRC-aided Polar codes. 
% Please note the differences in the code rates, especially in the case of $N \approx 250$.

\subsection{Second-Order Codes}

\begin{figure}[t]
    \centering
    \includegraphics[width=\linewidth]{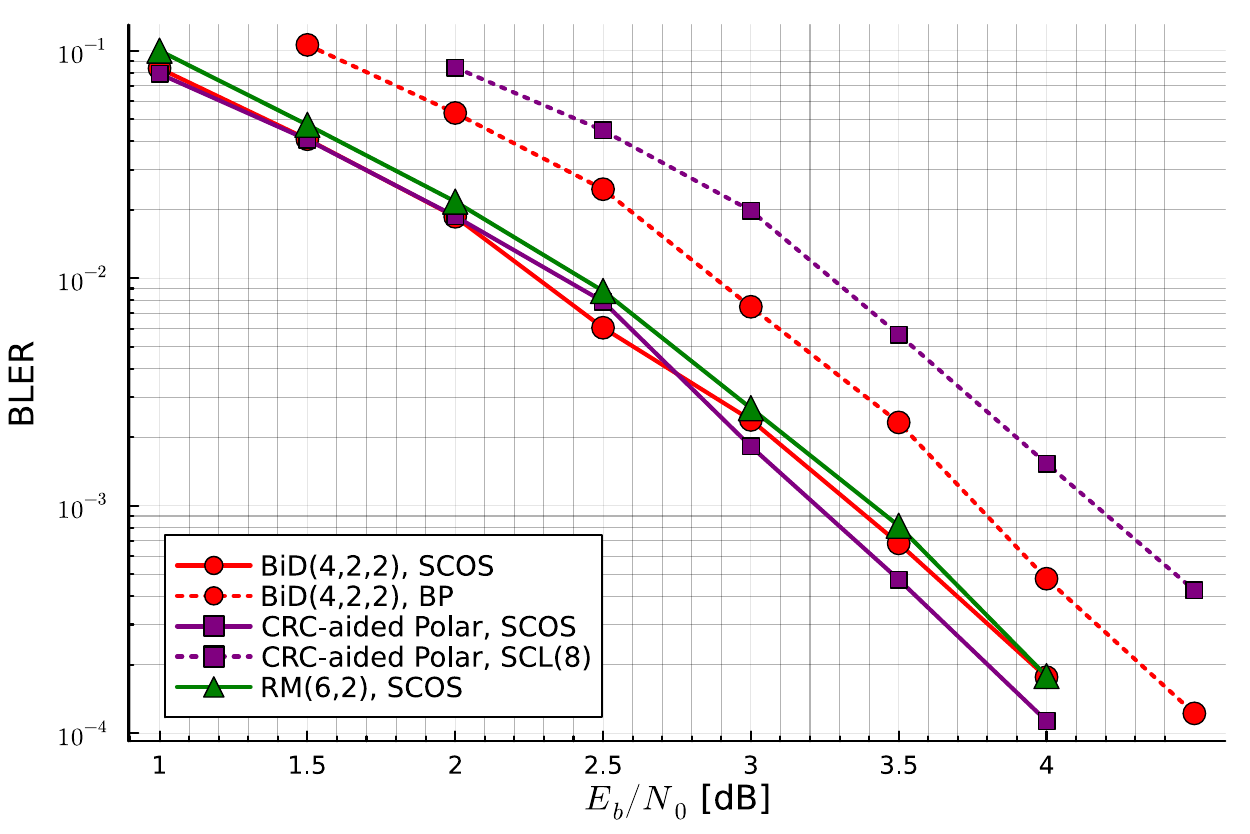}
    \caption{Second-order BiD code of length $81$.}
    \label{fig:81}
\end{figure}

\begin{figure}[t]
    \centering
    \includegraphics[width=\linewidth]{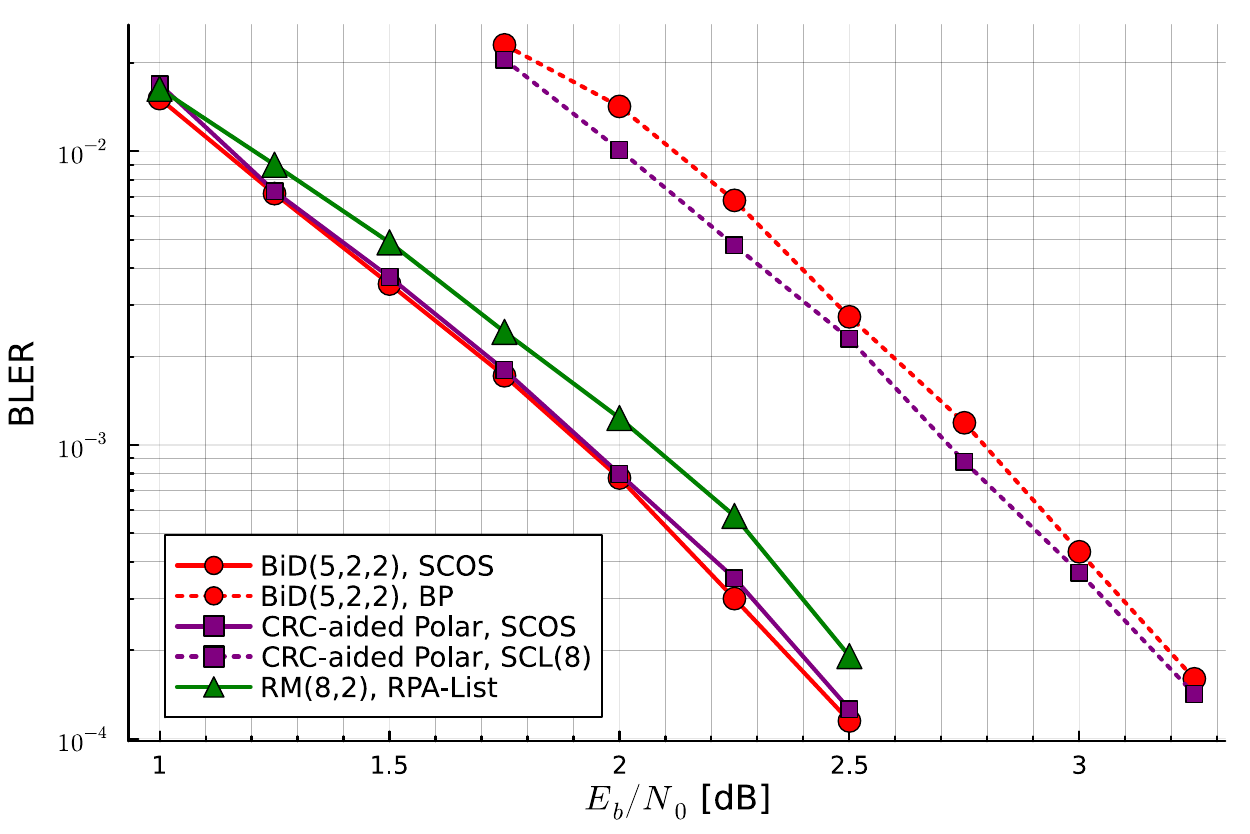}
    \caption{Second-order BiD code of length $243$.}
    \label{fig:243}
\end{figure}

We compare $\bid(4,2,2)$ ($N=81$, rate $R=0.296$) and $\bid(5,2,2)$ ($N=243$, $R=0.164$) with $\RM(6,2)$ ($N=64$,  $R=0.343$), $\RM(8,2)$ ($N=256$, $R=0.144$) and CRC-aided Polar codes in Fig.~\ref{fig:81} and~\ref{fig:243}. 
The CRC-aided Polar codes ($8$-bit CRC) in these figures are of lengths $64$ and $256$, respectively, with code rates equal to that of the corresponding BiD codes.
Each of these figures shows: \emph{(i)}~the BLERs of BiD code, RM code and CRC-aided Polar code, all of them under essentially-ML decoding; \emph{(ii)}~BiD code with the BP decoder from Section~\ref{sec:sub:BP_decoding}; and \emph{(iii)}~the CRC-aided Polar code with successive cancellation list (SCL) decoder with list size $8$.
We used SCOS~\cite{YuC_TCOM_24} to obtain the ML decoding performance of the BiD and CRC-aided Polar codes, while we used SCOS and RPA-list decoders~\cite{YeA_IT_20} for RM codes. 
We used the decoder library from~\cite{RBV_TVT_21,RBV_GitHub} for SCL decoding of CRC-aided Polar codes.
We observe that the BLER of the BiD codes under the BP decoder is within $1$~dB of the ML decoding performance and is similar to that of CRC-aided Polar codes under SCL($8$). 
The average number of BP iterations needed for convergence is small; for instance, this number is $2.3$ for $\bid(5,2,2)$ at $E_b/N_0=2$~dB.

Simulations for length $729$ show that the BP decoding of $\bid(6,2,2)$ performs about $0.25$~dB worse than SCL($8$) decoding of the CRC-aided Polar code ($11$-bit CRC) from 5G-NR (please see Appendix~\ref{app:fig:729}).

% \section{Discussion}

\subsection{Discussion}

Our BP decoder for second-order BiD codes has two deficiencies:~\emph{(i)}~the factor graph contains a large number of short cycles; and \emph{(ii)}~the factor graph corresponds to the super-code $\bid(m,2,2) \oplus \bid(m,0,0)$ instead of the second-order BiD code (see Appendix~\ref{app:bp_decoding}).
Incorporating additional low-weight parity checks and using techniques such as decimation~\cite{YLHAP_ISIT_24} might  bring the BLER closer to the ML decoding performance.

% The factor graph of second-order BiD codes contains a large number of short cycles. 
% Further, every codeword in $\abelian(m,\{0,2\})$ satisfies all the parity checks enforced in the factor graph. Thus, by design, this BP decoder decodes to $\abelian(m,\{0,2\})$ which is a super-code of $\bid(m,2,2)=\abelian(m,\{2\})$. 
% We believe that the performance of the BP decoder can be improved using intelligent counter-measures, such as smart message scheduling, decimation~\cite{YLHAP_ISIT_24}, and incorporating additional parity checks to ensure that only the codewords in $\bid(m,2,2)$ satisfy all the constraints.

% % % % % % % % % % % % % % % % % % % % 
% Bibliography
% \newpage 
\bibliographystyle{IEEEtran}
\IEEEtriggeratref{16}
\bibliography{IEEEabrv,ISIT26}

\newpage 
% % % % % % % % % % % % % % % % % % % % 
% Appendix

\appendix 

\subsection{Proof of Lemma~\ref{lem:automorphisms}} 
\label{app:lem:automorphisms}

We will argue that each of these three classes of permutations is a code automorphism.

\emph{Part~(i).} The permutation $X^{\p{i}} \to X^{\p{i} + \p{k}}$ sends a codeword $f = \sum_{\p{i}} f_{\p{i}} X^{\p{i}}$ to the polynomial $\sum_{\p{i}} f_{\p{i}} X^{\p{i} + \p{k}} = X^{\p{k}} f$. Since $\abelian$ is an ideal and $f \in \abelian$, we see that $X^{\p{k}} f \in \abelian$. 
% \textcolor{red}{$X^{\p{k}} f \in \abelian$} 
Hence, this permutation is an automorphism.

\emph{Part~(ii).} Suppose $f \in \abelian$ and $g$ is the image of $f$ under the permutation $X_{\ell} \to X_{\ell}^2$. We need to argue that $g \in \abelian$. Towards this, consider any $\p{j} \in \Zb_3^m$ with $\wt(\p{j}) \notin \Wc$. The evaluation of $g$ at the point $(\alpha^{j_1},\dots,\alpha^{j_m})$ is equal to the evaluation of $f$ at
\begin{equation} \label{eq:proof_of_lemma_automorphisms_1}
\left( \alpha^{j_1},\dots,\alpha^{j_{\ell-1}},\alpha^{2j_{\ell}},\alpha^{j_{\ell+1}},\dots,\alpha^{j_m} \right)
\end{equation}
since \mbox{$2^2=1$} in $\Zb_3$. Observe that 
\begin{equation*}
(j_1,\dots,j_{\ell-1},2j_{\ell},j_{\ell+1},\dots,j_m)
\end{equation*}
has the same Hamming weight as $\p{j}$ (since $2$ is a unit in $\Zb_3$), and hence, its weight does not lie in $\Wc$. Thus, the point~\eqref{eq:proof_of_lemma_automorphisms_1} is a root of $f$, and hence, $(\alpha^{j_1},\dots,\alpha^{j_m})$ is a root of $g$.

\emph{Part~(iii).} Similar to the proof of Part~\emph{(ii)}, suppose $g$ is the image of \mbox{$f \in \abelian$} under the permutation \mbox{$X_{\ell} \to X_{\gamma(\ell)}$} for all \mbox{$\ell \in \{1,\dots,m\}$}. Let \mbox{$\p{j} \in \Zb_3^m$} with \mbox{$\wt(\p{j}) \notin \Wc$}. The evaluation of $g$ at $(\alpha^{j_1},\dots,\alpha^{j_m})$ equals the evaluation of $f$ at $(\alpha^{j_{\gamma^{-1}(1)}},\dots,\alpha^{j_{\gamma^{-1}(m)}})$. Since $(j_{\gamma^{-1}(1)},\dots,j_{\gamma^{-1}(m)})$ has the same Hamming weight as $\p{j}$, its weight does not lie in $\Wc$, and hence, $f$ evaluates to zero at $(\alpha^{j_{\gamma^{-1}(1)}},\dots,\alpha^{j_{\gamma^{-1}(m)}})$.

% \subsection{Proof of Theorem~\ref{thm:projection}} \label{app:thm:projection}

% We will assume that $S=\{m-|S|+1,\dots,m\}$, because otherwise we can use an appropriate code automorphism (from part~(iii) of Lemma~\ref{lem:automorphisms}) to map $X_{s_1},\dots,X_{s_{|S|}}$ to $X_{m-|S|+1},\dots,X_{m}$, respectively. This maps the index components $i_{s_1},\dots,i_{s_{|S|}}$ to $i_{m-|S|+1},\dots,i_{m}$, respectively.

% Let us denote $\punc(\p{f},S,\p{u})$ by $\p{a}$ and $\punc(\p{f},S,\p{v})$ by $\p{b}$.
% Further, for any $\p{j} \in \Zb_3^m$, let us denote the first $m-|S|$ components of $\p{j}$ as $\p{j'}$ and the rest as $\p{j''}$.
% Using the same ideas as in the proofs of Lemmas~\ref{lem:puncturing} and~\ref{lem:projection} we have 
% \begin{equation*}
% \textstyle
% \hat{a}_{\p{j'}} + \hat{b}_{\p{j'}} = \sum_{\p{j''} \in \Zb_3^{|S|}} \hat{f}_{\p{j}} \left( \alpha^{-\p{u} \cdot \p{j''}} + \alpha^{-\p{v} \cdot \p{j''}} \right).
% \end{equation*}
% If $\wt(\p{j'}) \leq w-|S|-1$ or $\wt(\p{j'}) > w+1$, we have $\wt(\p{j}) \neq w$ for any choice of $\p{j''}$, and thus, $\hat{f}_{\p{j}}=0$, and consequently, $\hat{a}_{\p{j'}} + \hat{b}_{\p{j'}}=0$.
% If $\wt(\p{j'})=w$, then $\p{j''}=\p{0}$ is the only choice with possibly non-zero $\hat{f}_{\p{j}}$; in this case we have $\p{u} \cdot \p{j''} = \p{v} \cdot \p{j''} = 0$, and hence, $\hat{a}_{\p{j'}} + \hat{b}_{\p{j'}}=0$. Thus, we have proved that $\hat{a}_{\p{j'}} + \hat{b}_{\p{j'}} = 0$ if $\wt(\p{j'}) \notin \{w-|S|,\dots,w-1\}$.

\subsection{Proof of Lemma~\ref{lem:bidd_lower_bound_5}} \label{app:lem:bidd_lower_bound_5}

We will use $\spanof$ to denote the linear span of a collection of vectors. 
% We will use the fact that $\Zb_3^{w-1}$ is a vector space over the finite field $\Zb_3$.
% and $\rowspaceof$ to denote the span of the rows of a matrix.
% 
We already know that no codewords of weight $1$ exist in $\bidd(m,2,2)$.
It is enough to show that $\mathscr{M}$ is empty for $w=2,3,4$. 
% \subsection{Non-Existence Proofs for $w \le 4$}
% \subsubsection{}
% For $w=1$, the sum is empty and $0=1$ is impossible.

\emph{Case~(i): $w=2$.}
In this case $S = \{0\}$ and $\p{M}$ contains only one row. 
We use the fact that the linear span of $V$ over $\Zb_3$ equals $\Zb_3^m$. This can be proved, in a straightforward way, by showing that every standard basis vector of $\Zb_3^m$ lies in the span of $V$. For instance, 
% $(1,0,\dots,0)$ is the sum of $(2,0,\dots,0,2)$, $(2,1,0,\dots,0)$ and the $(m-2)$ right-cyclic shifts of the latter vector.  
{$(1,0,\dots,0)$ is the sum of $(2,2,0,\dots,0)$ and $(2,1,0,\dots,0)$}.
Thus, if $\p{j}\p{M}^T = \p{0}$ for all weight-$2$ vectors $\p{j}$, the only solution is $\p{M} = \p{0}$. This contradicts the requirement that the rows of $\p{M}$ are non-zero. Hence $\mathscr{M}$ is empty.

\emph{Case~(ii): $w=3$.}
Here $S = \{(1, 2), (2, 1)\}$. %note that $S \cup \{\p{0}\}$ is a line in $\Zb_3^2$. 
% Recall that Lemma~\ref{lem:min_wt_reformulation} imposes constraints on the columns $\p{c}_1^T,\dots,\p{c}_m^T$ of $\p{M}$. 
% From~\eqref{eq:min_wt_reformulation_columns} and the fact $\p{0} \notin S$ we deduce that the columns of $\p{M}$ are distinct.
% Considering those $\p{j} \in V$ whose two non-zero entries are $1$ and $2$ and the fact $\p{0} \notin S$, we deduce that all columns of $\p{M}$ must be distinct.
% Denoting the $i^\tth$ column of $\p{M}$ by $\p{c}^T_i$ we have $\p{c}_i \pm \p{c}_j \in S$.
Considering the first three columns of $\p{M}$, from~\eqref{eq:min_wt_reformulation_columns}, we have $\p{c}_1-\p{c}_2,\p{c}_1-\p{c}_3 \in S$. If these two differences are equal, we have $\p{c}_2=\p{c}_3$ which implies, via~\eqref{eq:min_wt_reformulation_columns}, $\p{0} = \p{c}_2 - \p{c}_3 \in S$, a contradiction. 
On the other hand, if $\p{c}_1-\p{c}_2 \neq \p{c}_1-\p{c}_3$, then these two vectors must be negatives of each other (since the two vectors in $S$ are negatives of each other), and hence, $\p{c}_2+\p{c}_3 = 2\p{c}_1$. 
Thus we have $S = \{\p{c}_1,2\p{c}_1\}$.
% The condition $\p{c}_2+\p{c}_3,\p{c}_2-\p{c}_3 \in \{\p{c}_1,2\p{c}_1\}$ implies that at least one of the following must be $\p{0}$: $\p{c}_1-\p{c}_3$, $\p{c}_1+\p{c}_3$, $\p{c}_2-\p{c}_3$, $\p{c}_2+\p{c}_1$. From~\eqref{eq:min_wt_reformulation_columns} and the fact $\p{0} \notin S$, this is a contradiction. 
{The condition $\p{c}_2+\p{c}_3,\p{c}_2-\p{c}_3 \in \{\p{c}_1,2\p{c}_1\}$ forces either $\p{c}_2$ or $\p{c}_3$ to be $\p{0}$. This implies that either $2\p{c}_1 - \p{c}_3$ or $2\p{c}_1 - \p{c}_2$ must be $\p{0}$. From~\eqref{eq:min_wt_reformulation_columns} and the fact $\p{0} \notin S$, this is a contradiction.} 
% $S \cup \{\p{0}\}$ is a subspace and $\p{c}_2,\p{c}_3 \neq \p{0}$, we deduce that $\p{c}_2,\p{c}_3 \in S$. Thus, $S$, which is of size $2$, contains three distinct vectors $\p{c}_1,\p{c}_2,\p{c}_3$. This is a contradiction.
% For any two columns $\p{c}_i^T$ and $\p{c}_j^T$, we have $\p{c}_i - \p{c}_j \in S$ and $\p{c}_i + \p{c}_j \in S$. This implies $c_2 = c_1 - t$ and $c_3 = c_1 - 2t$ for $t \in S$. The condition $c_2 + c_3 \in S$ forces $c_1 \in S$ as well as $c_2,c_3 \in S$. However, as $|S|=2$, it is impossible to select three distinct columns $c_1, c_2, c_3 \in S$.

\emph{Case~(iii): \mbox{$w=4$}.}
Here $S$ consists of all permutations of $(k, k, 0)$ for $k \in \{0, 1, 2\}$. This is the union of three linearly independent lines in $\Zb_3^3$, viz., $S_1 = \spanof(1,1,0)$, $S_2 = \spanof(1,0,1)$, and $S_3 = \spanof(0,1,1)$. That is, $S = S_1 \cup S_2 \cup S_3$. 

Since the rows of $\p{M}$ are distinct, at least $2$ columns of $\p{M}$ must be distinct.

\emph{Sub-case 1: Only two columns are distinct.}
Let $\p{c}_1 = \p{c}_2 \neq \p{c}_3$. We have $2\p{c}_1 + 2\p{c}_2 = 4\p{c}_1 = \p{c}_1 \in S$. Similarly, $\p{c}_1 + \p{c}_2 = 2\p{c}_1 \in S$. Let $\p{c}_1 + \p{c}_3 = \p{t}_a \in S$ and $\p{c}_1 - \p{c}_3 = \p{t}_b \in S$. Then $\p{t}_a + \p{t}_b = 2\p{c}_1 \in S$. The sum of two vectors in $S$ belongs to $S$ only if they lie on the same line. Thus, $\p{t}_a, \p{t}_b$ and $\p{c}_1$ belong to the same line. Their difference $\p{t}_a - \p{t}_b = 2\p{c}_3$ forces $\p{c}_3$ to lie on that same line as well.
Consequently, every column of $\p{M}$ lies on the same line.

\emph{Sub-case 2: $\p{M}$ has at least three distinct columns.}
Assume $\p{c}_1, \p{c}_2, \p{c}_3$ are distinct. Consider $\p{t}_1 = \p{c}_1 - \p{c}_2 \in S$ and $\p{t}_2 = \p{c}_1 - \p{c}_3 \in S$. Since their difference $(\p{c}_1 - \p{c}_2) - (\p{c}_1 - \p{c}_3) = \p{c}_3 - \p{c}_2$ belongs to $S$, $\p{t}_1$ and $\p{t}_2$ must be linearly dependent and $\p{t}_2 = 2\p{t}_1$. We then have $\p{c}_2 = \p{c}_1 - \p{t}_1$ and $\p{c}_3 = \p{c}_1 + \p{t}_1$. Then the condition $\p{c}_2 + \p{c}_3 \in S$ simplifies to
% \begin{equation*}
%     (c_1 - t_1) + (c_1 - 2t_1) = 2c_1 - 3t_1 = 2c_1 \in S
% \end{equation*}
$2\p{c}_1 \in S$ which implies $\p{c}_1 \in S$ since $S$ is a union of lines.
% Since $2c_1 \in S$, it follows that $c_1 \in S$. 
Note that $S$ contains $\p{c}_1+\p{c}_2 = 2\p{c}_1+2\p{t}_1=2\p{c}_3$, hence, $\p{c}_3 \in S$. Thus, we have $\p{c}_1,\p{c}_3,\p{c}_1-\p{c}_3 \in S$. This implies that $\p{c}_1$ and $\p{c}_3$ must lie on the same line.
From the symmetry of the problem we conclude that $\p{c}_1,\p{c}_2,\p{c}_3$ (and all subsequent columns of $\p{M}$) lie on the same line.
% The simultaneous conditions $c_1 \in S$ and $(c_1 - c_2) = t_1 \in S$ force $c_1$ to lie on the same line as $t_1$. Consequently, $c_2, c_3$ and all other subsequent columns fall on the same line.

In both sub-cases, there exists an $i$ such that all columns of $\p{M}$ belong to $S_i$. Since every $S_i$ contains a zero at a fixed coordinate, the corresponding row in $\p{M}$ is $\p{0}$. This contradicts the requirement that the rows of $\p{M}$ be non-zero. Thus, weight-$4$ codewords do not exist.    

\subsection{Polynomial Form of Minimum-Weight Codewords of $\bidd(m,2,2)$, $m \geq 4$} \label{app:polynomial_form_wt_6_codewords}

Using the automorphisms identified in Lemma~\ref{lem:automorphisms} on $f$ from~\eqref{eq:weight_6_polynomial}, we enumerate the distinct resulting polynomials as follows:
\begin{itemize}
\item Variable permutations (type~\emph{(iii)}): Since $f$ is symmetric in $X_2, \dots, X_m$, applying these permutations gives $m$ distinct polynomials.
\item Replacing $X_{\ell}$ with $X_{\ell}^2$ (type~\emph{(ii)}): Note that the transformation $X_1 \to X_1^2$ preserves $f$. Additionally, for the remaining variables $\{X_2, \dots, X_m\}$, squaring a specific subset is equivalent to squaring the remaining variables in the set because $X_{\ell}^3=1$ in $\Rm$. This symmetry results in $2^{m-2}$ distinct polynomials.
\item Translations (type~\emph{(i)}): Shifting by $\p{k} \in \Zb^m_3$ potentially gives $3^m$ variations. However, due to the symmetry $f = X_2\dots X_mf = X_2^2\dots X_m^2f$, we obtain only $3^{m-1}$ distinct polynomials.
\end{itemize}

Using these three results, we obtain $m2^{m-2}3^{m-1}$ distinct polynomials of weight $w=6$ for $\bidd(m,2,2)$. This matches exactly the number of codewords found using the computer search (Theorem~\ref{thm:bidd_dmin}).

\subsection{Span of Minimum-Weight Codewords of $\bidd(m,2,2)$, $m \geq 4$} \label{app:span_of_min_wt_codewords}

For a given $m \geq 4$, let $\Cs_{\min}$ denote the $\Fb_2$-span of all the minimum-weight codewords of $\bidd(m,2,2)$. 
We will first argue that $\Cs_{\min}$ is an ideal in $\Rm$.
From Lemma~\ref{lem:automorphisms} we know that for any $\p{k} \in \Zb_3^m$ and $f \in \bidd(m,2,2)$ the product $X^{\p{k}}f \in \bidd(m,2,2)$ and it has the same weight as $f$. Thus, if $f$ is a minimum-weight codeword of $\bidd(m,2,2)$ then so is $X^{\p{k}} f$. We deduce that for any $g = \sum_{\p{k} \in \Zb_3^m} g_{\p{k}} X^{\p{k}} \in \Rm$ and any minimum-weight codeword $f \in \bidd(m,2,2)$, the product $gf$ is a linear combination of minimum-weight codewords, and hence, $gf \in \Cs_{\min}$. As an immediate consequence of this observation we see that for any $g \in \Rm$ and $f \in \Cs_{\min}$ we have $gf \in \Cs_{\min}$. Thus, $\Cs_{\min}$ is an ideal.

We note that $\Rm$ is the group algebra of the abelian group $(\Zb_3^m,+)$ over the field $\Fb_2$. Since the characteristic of the field $\Fb_2$ does not divide the order $3^m$ of this group, $\Rm$ is a semisimple ring and its ideals are direct sums of its minimal ideals~\cite{Ber_Cybernetics_II_67,MacWilliams_Bell_70}. 
The ideals of $\Rm$ can be characterized via the DFT and the conjugacy classes~\cite{RaS_IT_92}. 
We partition $\Zb_3^m$ into $1 + (3^m -1)/2$ conjugacy classes, where each of these classes is a minimal set closed under multiplication by $2$. These are $C_{\p{0}} = \{\p{0}\}$ and $C_{\p{j}} = \{\p{j},2\p{j}\}$ for non-zero $\p{j}$. 
Let $\mathscr{F}$ be the family of all conjugacy classes; we have $|\mathscr{F}| = 1+(3^m-1)/2$ and $\cup_{C \in \mathscr{F}} \, C = \Zb_3^m$.
The ideals $\mathscr{I} \subseteq \Rm$ are in a one-to-one correspondence with the subsets $S_{\mathscr{I}} \subseteq \mathscr{F}$. Thus, the number of ideals in $\Rm$ is $2^{|\mathscr{F}|}$. 
The points in $\Zb_3^m$ covered by the conjugacy classes in $S_{\mathscr{I}}$ denote the common zeros of all $f \in \mathscr{I}$ when evaluated at $(\Fb_4^*)^m$, that is,
% The collection of conjugacy classes $S_{\mathscr{I}}$ associated with the ideal $\mathscr{I}$ satisfies the following property
\begin{equation*}
\{ \p{j} \in \Zb_3^m : \hat{f}_{\p{j}} = 0 ~\forall~f \in \mathscr{I} \} = \bigcup_{C \in S_{\mathscr{I}}} C.
\end{equation*}
We now identify the ideal $\Cs_{\min}$ by identifying the corresponding subset $S_{\Cs_{\min}} \subseteq \mathscr{F}$.

We first show that $S_{\Cs_{\min}}$ includes all conjugacy classes $C_{\p{j}}$ with \mbox{$\wt(\p{j})=0,2$}. The fact that $\Cs_{\min}$ is spanned by a collection of codewords from $\bidd(m,2,2)$ implies that for every $f \in \Cs_{\min}$, we have $\hat{f}_{\p{j}}=0$ for all $\p{j}$ with $\wt(\p{j})=2$. Further, the polynomial in~\eqref{eq:weight_6_polynomial} and any other polynomial obtained from~\eqref{eq:weight_6_polynomial} by applying the automorphisms from Lemma~\ref{lem:automorphisms} vanish at the point $(1,\dots,1)=(\alpha^0,\dots,\alpha^0)$. Since these are exactly all the minimum-weight codewords of $\bidd(m,2,2)$ (see Appendix~\ref{app:polynomial_form_wt_6_codewords}), we deduce that $S_{\Cs_{\min}}$ includes the conjugacy class $C_{\p{0}}$ as well.
% all conjugacy classes $C_{\p{j}}$ with $\wt(\p{j}) \in \{0,2\}$. 

We next show that any $C_{\p{j}}$ with $\wt(\p{j}) \neq 0,2$ is not included in $S_{\Cs_{\min}}$. We prove this by showing that for any $\p{j}$ with weight other than $0$ or $2$, there exists a minimum-weight codeword $f \in \bidd(m,2,2)$ such that $\hat{f}_{\p{j}} \neq 0$. 
This immediately characterizes $\Cs_{\min}$ as 
\begin{equation*}
\Cs_{\min} = \left\{ f \in \Rm : \hat{f}_{\p{j}} = 0 ~\forall~\p{j} \text{ with } \wt(\p{j})=0,2 \right\}.
\end{equation*}
That is, $\Cs_{\min}$ is the abelian code $\abelian(m,\{0,\dots,m\} \setminus \{0,2\})$. 
Hence, the dual code of $\Cs_{\min}$ is 
\begin{align*}
\Cs_{\min}^\perp &= \abelian(m,\{0,2\}) \\
&= \abelian(m,\{2\}) \oplus \abelian(m,\{0\}) \\
&= \bid(m,2,2) \oplus \bid(m,0,0).
\end{align*}
% $\abelian(m,\{0,2\}) =  \abelian(m,\{2\}) \cup \abelian(m,\{0\}) = \bid(m,2,2) \oplus \bid(m,0,0)$. 
It is straightforward to verify that $\bid(m,0,0)$ is the repetition code. 
We complete this subsection by proving 

\begin{lemma}
If $m \geq 4$ and $\p{j} \in \Zb_3^m$ with $\wt(\p{j}) \neq 0,2$ then there exists a minimum-weight codeword $f$ of $\bidd(m,2,2)$ such that $\hat{f}_{\p{j}} \neq 0$.
\end{lemma}
\begin{IEEEproof}
For a given $\p{j}$ we identify a suitable $f$ with $X^{\p{0}}$ as one of its terms using the framework developed in Section~\ref{sec:sub:min_wt_parity_checks}. 
Note that such an $f$ corresponds to a $5 \times m$ matrix $\p{M}$ whose columns are picked from one of the cliques $C_1$ or $C_2$ in~\eqref{eq:cliques_C1_C2}. The value of DFT at $\p{j}$ is non-zero if and only if $\p{j} \p{M}^T \notin S$, where $S$ is the subset of $\Zb_3^{5}$ defined in~\eqref{eq:S_definition} (with $w=6$). 
Recall that $S$ consists of all permutations of the vector $(0,a,a,b,b)$ for $a,b \in \{0,1,2\}$.

To illustrate the construction of $\p{M}$ for a given $\p{j}$ we consider $\p{j}$ of the form $(1,1,\dots,1,0,0,\dots,0)$, i.e., the non-zero entries of $\p{j}$ are all equal to $1$ and they appear on the left part of $\p{j}$. The proofs for other choices of $\p{j}$ are similar and can be obtained by using the automorphisms in Lemma~\ref{lem:automorphisms} with the construction provided here.
We will use the clique $C_1$ in our construction of $\p{M}$, and we will denote the vectors in $C_1$ as $\p{v}_1 = (0,0,1,1,1)$, $\p{v}_2=(1,2,0,1,2)$ and $\p{v}_3=(2,1,0,2,1)$. 

\emph{Case~{(i)}. $\wt(\p{j})=1$:} In this case we can choose any $\p{M}$ whose first column is $\p{v}_1^T$. Then $\p{j}\p{M}^T=\p{v}_1 \notin S$.

\emph{Case~{(ii)}. $\wt(\p{j})=t \geq 3$ and $t$ is odd.} We choose $\p{v}_1^T$ as the first column of $\p{M}$, and the remaining columns are alternated between $\p{v}_2^T$ and $\p{v}_3^T$, that is, the columns of $\p{M}$ are $\p{v}_1^T,\p{v}_2^T,\p{v}_3^T,\p{v}_2^T,\p{v}_3^T,\dots$ in that order. Since $\p{v}_2+\p{v}_3 = \p{0}$, we have $\p{j}\p{M}^T=\p{v}_1 \notin S$.

\emph{Case~{(iii)}. $\wt(\p{j})=t \geq 4$ and $t$ is even.} Here we choose $\p{v}_1^T$ as the first column of $\p{M}$, $\p{v}_2^T$ as the next three columns of $\p{M}$, and then alternate between $\p{v}_2^T$ and $\p{v}_3^T$. Since $3\p{v}_2=\p{0}$ and $\p{v}_2 + \p{v}_3=0$, we have $\p{j}\p{M}^T = \p{v}_1 \notin S$.
\end{IEEEproof}

\subsection{On the Dual Distance of General BiD Codes} \label{app:dual_distance_recursion}

% In~\cite[Theorem~3.1]{DNNK_ITW_25} a recursive algorithm was presented to numerically compute upper and lower bounds on the minimum distance of an arbitrary BiD code. The key ideas used in this technique are: 
% \emph{(i)}~the recursive formulation~\cite[Lemma~2.3]{DNNK_ITW_25} of the generator matrix of $\abelian(m,\{w\})$ in terms of $\abelian(m-1,\{w\})$ and $\abelian(m-1,\{w-1\})$; and
% \emph{(ii)}~partitioning a codeword $\p{\rho}$ into three sub-vectors $\p{\rho}_0,\p{\rho}_1,\p{\rho}_2$ and relating $\wt(\p{\rho})$ to the weights of $\p{\rho}_i$, $i=0,1,2$.
% Fortuitously, this technique from~\cite{DNNK_ITW_25} works just as it is for any abelian code $\abelian(m,\Wc)$ since it does not rely on the peculiarities of BiD codes beyond the fact that~\cite[Lemma~2.3]{DNNK_ITW_25} applies to them.

% In this section we first revisit the recursive technique of~\cite{DNNK_ITW_25} in the context of applying the idea to any abelian code $\abelian(m,\Wc)$ and not just BiD codes, and then present the numerical results for the specific case of duals of BiD codes. 

We first review the technique from~\cite{DNNK_ITW_25} (which was described for BiD codes only) as applied to any abelian code $\abelian(m,\Wc)$.
Since the proof of~\cite[Theorem~3.1]{DNNK_ITW_25} applies almost verbatim to this more general setting (with only straightforward modifications to account for the larger family of codes), we keep our presentation as concise as possible.

\subsubsection{Recursive Bounds on Minimum Distance of $\abelian(m,\Wc)$.}

Let $\otimes$ denote the Kronecker product of matrices. We use $\left[\,\p{a}_1; \, \p{a}_2; \, \cdots ; \, \p{a}_\ell\,\right]$ to denote the matrix with rows $\p{a}_1,\dots,\p{a}_{\ell}$.
Recall from~\cite[Lemma~2.3]{DNNK_ITW_25} that the generator matrix $\p{G}_{m,w}$ of $\abelian(m,\{w\})$ satisfies the recursion
\begin{align} \label{eq:G_mw_matrix_recursion}
\p{G}_{m,w} = 
\left[ \!\!
\begin{array}{l}
(1,1,1) \otimes \p{G}_{m-1,w} \\
(1,1,0) \otimes \p{G}_{m-1,w-1} \\
(1,0,1) \otimes \p{G}_{m-1,w-1}
\end{array}
\!\! \right]
\text{ for } 1 \leq w \leq m-1, 
\end{align}
$\p{G}_{m,0} = [1~1~1]^{\otimes m} = [1~1~1] \otimes \p{G}_{m-1,0}$, and finally, $\p{G}_{m,m}=[1~1~0;~1~0~1]^{\otimes m} = [1~1~0;~1~0~1] \otimes \p{G}_{m-1,m-1}$.
Denote the generator matrix of $\abelian(m,\Wc)$ as $\p{G}_{m,\Wc}$, which (from the direct-sum decomposition property) is the vertical concatenation of $\p{G}_{m,w}$, $w \in \Wc$.
For a given $(m,\Wc)$, we define 
\begin{align*}
\Wc_x &\triangleq \left\{ w \in \Wc : w \neq m \right\},  \\
\Wc_y &\triangleq \left\{w-1 : w \in \Wc, w \neq 0 \right\}.
\end{align*}
It follows from~\eqref{eq:G_mw_matrix_recursion} that 
\begin{equation*}
\p{G}_{m,\Wc} = 
\left[ \!
\begin{array}{l}
(1,1,1) \otimes \p{G}_{m-1,\Wc_x} \\
(1,1,0) \otimes \p{G}_{m-1,\Wc_y} \\
(1,0,1) \otimes \p{G}_{m-1,\Wc_y}
\end{array}
\! \right].
\end{equation*}
Hence, for every $\p{\rho} \in \abelian(m,\Wc)$ there exist unique 
\begin{equation*}
\p{a} \in \abelian(m-1,\Wc_x) \text{ and } \p{a}',\p{a}'' \in \abelian(m-1,\Wc_y)    
\end{equation*}
such that 
\begin{equation*} % \label{eq:rho_kronecker_decomp}
\p{\rho} = (1,1,1) \otimes \p{a} + (1,1,0) \otimes \p{a}' + (1,0,1) \otimes \p{a}''.   
\end{equation*}
Further, we have 
\begin{equation*}
\textstyle
\p{a} = \sum_{w \in \Wc_x} \p{a}_w, ~\p{a}' = \sum_{w \in \Wc_y} \p{a}_w',~\text{ and }~ \p{a}''=\sum_{w \in \Wc_y} \p{a}_w''
\end{equation*}
for some choice of $\p{a}_w,\p{a}_w',\p{a}_w'' \in \abelian(m-1,\{w\})$.
Splitting $\p{\rho}$ as $(\p{\rho}_0,\p{\rho}_1,\p{\rho}_2)$ we have 
\begin{align} \label{eq:rho_012}
\begin{split}
\textstyle 
\p{\rho}_0 &= \sum_{w \in \Wc_x} \p{a}_w + \sum_{w \in \Wc_y} \left( \p{a}_w' + \p{a}_w'' \right) \\
\p{\rho}_1 &= \sum_{w \in \Wc_x} \p{a}_w + \sum_{w \in \Wc_y} \p{a}_w'  \text{ and} \\ 
\p{\rho}_2 &= \sum_{w \in \Wc_x} \p{a}_w + \sum_{w \in \Wc_y} \p{a}_w''. 
\end{split}
\end{align} 

We will use $d_{m}(\Wc)$ to denote the minimum distance of $\abelian(m,\Wc)$.
We will analyze the minimum non-zero weight of $\p{\rho}$ under four distinct cases.

\emph{Case 1: $\p{\rho}_0=\p{\rho}_1=\p{\rho}_2=\p{0}$.} 
This corresponds to $\p{\rho}=\p{0}$, which is not useful.

\emph{Case 2: exactly one among $\p{\rho}_0,\p{\rho}_1,\p{\rho}_2$ is non-zero.} 
Assume $\p{\rho}_0 \neq \p{0}$; the analysis of the other two possibilities is similar. 
The fact $\p{\rho}_1=\p{\rho}_2=\p{0}$ implies 
\begin{align*}
&\p{a}_w = \p{a}_w' = \p{a}_w'' \text{ if } w \in \Wc_x \cap \Wc_y, \text{ and } \\
&\p{a}_w = \p{a}_w' = \p{a}_w'' = \p{0} \text{ otherwise}.
\end{align*}
Thus, $\p{\rho}_0 = \sum_{w \in \Wc_x \cap \Wc_y} \p{a}_w \in \abelian(m-1,\Wc_x \cap \Wc_y)$.
Hence, the smallest weight of $\p{\rho}$ under Case~2 is 
\begin{equation*}
D_2 \triangleq d_{m-1}\left( \Wc_x \cap \Wc_y\right).
\end{equation*}
In case $\Wc_x \cap \Wc_y$ is empty, $\abelian(m-1,\Wc_x \cap \Wc_y)$ is the trivial code $\{\p{0}\}$. No non-zero codewords exist in this code. In such a case we set $D_2 = \infty$.

\emph{Case 3: exactly two among $\p{\rho}_0,\p{\rho}_1,\p{\rho}_2$ are non-zero.}
Consider \mbox{$\p{\rho}_0=\p{0}$}; analysis of the other two possibilities is similar).
From~\eqref{eq:rho_012}, we see that $\p{a}_w = \p{a}'_w + \p{a}_w''$ for all $w \in \Wc_x \cap \Wc_y$ and $\p{a}_w=\p{a}_w'+\p{a}_w''=\p{0}$ otherwise. Hence we have 
\begin{equation*}
\p{\rho}_1 = \sum_{w \in \Wc_y} \!\!\! \p{a}_w'' ~\text{ and }~ \p{\rho}_2 = \sum_{w \in \Wc_y} \!\!\!  \p{a}_w'.
\end{equation*}
The smallest non-zero weight of $\p{\rho}$ in this case is
\begin{equation*}
D_3 \triangleq 2 d_{m-1}(\Wc_y).
\end{equation*}

\emph{Case~4: $\p{\rho}_0,\p{\rho}_1,\p{\rho}_2 \neq \p{0}$.} 
We identify two independent lower bounds for this case (and utilize their maximum) and one upper bound on $d_m(\Wc)$.

\begin{enumerate}
\item[\emph{4a.}] Since each $\p{\rho}_i \in \abelian(m-1,\Wc_x \cup \Wc_y)$, we have
\begin{equation*}
\wt\left( \p{\rho} \right) \geq D_{4a} \triangleq 3d_{m-1}(\Wc_x \cup \Wc_y)    .
\end{equation*}

\item[\emph{4b.}] For the alternative lower bound, consider two subcases: 
\emph{(i)}~$\p{\rho}_0,\p{\rho}_1,\p{\rho}_2$ are all equal, and 
\emph{(ii)}~at least two of them are unequal. 

Under subcase~\emph{(i)}, we have $\p{a}_w'=\p{a}_w'' = \p{0}$ for all $w \in \Wc_y$ and $\p{\rho}_i = \sum_{w \in \Wc_x} \p{a}_w$ for all $i$. The smallest value $\wt(\p{\rho})$ in this subcase is
\begin{equation*}
D_4' \triangleq 3d_{m-1}(\Wc_x).
\end{equation*}
For subcase~\emph{(ii)}, assume $\p{\rho}_1 \neq \p{\rho}_2$ (other possibilities can be handled in a similar way). From triangle inequality,
\begin{equation*}
\wt(\p{\rho}) = \sum_{i=0}^{2} \wt(\p{\rho}_i) \geq \wt(\p{\rho}_0) + \wt(\p{\rho}_1 + \p{\rho}_2).
\end{equation*}
Since 
\begin{equation*}
\p{\rho}_0 \in \abelian(m-1,\Wc_x \cup \Wc_y) \text{ and } \p{\rho}_1 + \p{\rho}_2 \in \abelian(m-1,\Wc_y),    
\end{equation*}
we see that 
\begin{equation*}
\wt(\p{\rho}) \geq d_{m-1}(\Wc_x \cup \Wc_y) + d_{m-1}(\Wc_y).
\end{equation*}
\end{enumerate}
In summary, under Case~4, the smallest value of $\wt(\p{\rho})$ is upper bounded by $D_4'$
% \begin{equation*}
% D'_4 \triangleq 3d_{m-1}(\Wc)
% \end{equation*}
and lower bounded by 
\begin{equation*}
D_4 \triangleq \max \! \Big\{\! D_{4a}, \, \min\{D'_4,d_{m-1}(\Wc_x \cup \Wc_y)+d_{m-1}(\Wc_y)\} \Big\}.
\end{equation*}

Gathering all the above observations, we arrive at the main result of this subsection.

\begin{theorem} \label{thm:dmin_recursion_bounds}
For any $m \geq 2$ and $\Wc \subseteq \{0,\dots,m\}$, with $D_2,D_3,D_4$ and $D'_4$ as defined above, we have
\begin{align*} % \label{eq:dmin_recursion_bounds}
\min\{D_2,D_3,D_4\} \leq d_m(\Wc) \leq \min\{D_2,D_3,D'_4\}.
\end{align*}
\end{theorem}

% % % table of minimum distances 
\renewcommand{\arraystretch}{1.10}
\begin{table}[htbp]
\centering
\caption{All Dual BiD Codes of Lengths $9$ to $243$}
\label{table:dual_bid_9_to_243}
\begin{tabular}{|c|c|c|c|c|}
\hline
\hline
$m$ & $r_1$ & $r_2$ & $\dmin$& $K$ \\
\hline
\hline
% 0 & 0 & 0 & 1 & 1 \\ \hline
% 1 & 0 & 0 & 3 & 1 \\ \hline
% 1 & 0 & 1 & 1 & 3 \\ \hline
% 1 & 1 & 1 & 2 & 2 \\ \hline
\multicolumn{5}{c}{Codes of length $9$} \\ \hline 
$2$ & $0$ & $0$ & $2$ & $8$\\ \hline
$2$ & $0$ & $1$ & $4$ & $4$\\ \hline
$2$ & $1$ & $1$ & $3$ & $5$\\ \hline
$2$ & $1$ & $2$ & $9$ & $1$\\ \hline
$2$ & $2$ & $2$ & $3$ & $5$\\ \hline
\hline 
\multicolumn{5}{c}{Codes of length $27$} \\ \hline 
$3$ & $0$ & $0$ & $2$ & $26$\\ \hline
$3$ & $0$ & $1$ & $4$ & $20$\\ \hline
$3$ & $0$ & $2$ & $8$ & $8$\\ \hline
$3$ & $1$ & $1$ & $3$ & $21$\\ \hline
$3$ & $1$ & $2$ & $8$ & $9$\\ \hline
$3$ & $1$ & $3$ & $27$ & $1$\\ \hline
$3$ & $2$ & $2$ & $5$ & $15$\\ \hline
$3$ & $2$ & $3$ & $9$ & $7$\\ \hline
$3$ & $3$ & $3$ & $3$ & $19$\\ \hline
\hline 
\multicolumn{5}{c}{Codes of length $81$} \\ \hline 
$4$ & $0$ & $0$ & $2$ & $80$\\ \hline
$4$ & $0$ & $1$ & $4$ & $72$\\ \hline
$4$ & $0$ & $2$ & $8$ & $48$\\ \hline
$4$ & $0$ & $3$ & $16$ & $16$\\ \hline
$4$ & $1$ & $1$ & $3$ & $73$\\ \hline
$4$ & $1$ & $2$ & $8$ & $49$\\ \hline
$4$ & $1$ & $3$ & $16$ & $17$\\ \hline
$4$ & $1$ & $4$ & $81$ & $1$\\ \hline
$4$ & $2$ & $2$ & $6$ & $57$\\ \hline
$4$ & $2$ & $3$ & $15$-$16$ & $25$\\ \hline
$4$ & $2$ & $4$ & $27$ & $9$\\ \hline
$4$ & $3$ & $3$ & $6$-$9$ & $49$\\ \hline
$4$ & $3$ & $4$ & $9$ & $33$\\ \hline
$4$ & $4$ & $4$ & $3$ & $65$\\ \hline
\hline 
\multicolumn{5}{c}{Codes of length $243$} \\ \hline 
$5$ & $0$ & $0$ & $2$ & $242$\\ \hline
$5$ & $0$ & $1$ & $4$ & $232$\\ \hline
$5$ & $0$ & $2$ & $8$ & $192$\\ \hline
$5$ & $0$ & $3$ & $16$ & $112$\\ \hline
$5$ & $0$ & $4$ & $32$ & $32$\\ \hline
$5$ & $1$ & $1$ & $3$ & $233$\\ \hline
$5$ & $1$ & $2$ & $8$ & $193$\\ \hline
$5$ & $1$ & $3$ & $16$ & $113$\\ \hline
$5$ & $1$ & $4$ & $32$ & $33$\\ \hline
$5$ & $1$ & $5$ & $243$ & $1$\\ \hline
$5$ & $2$ & $2$ & $6$ & $203$\\ \hline
$5$ & $2$ & $3$ & $16$ & $123$\\ \hline
$5$ & $2$ & $4$ & $32$ & $43$\\ \hline
$5$ & $2$ & $5$ & $81$ & $11$\\ \hline
$5$ & $3$ & $3$ & $7$-$12$ & $163$\\ \hline
$5$ & $3$ & $4$ & $21$-$27$ & $83$\\ \hline
$5$ & $3$ & $5$ & $27$ & $51$\\ \hline
$5$ & $4$ & $4$ & $7$-$9$ & $163$\\ \hline
$5$ & $4$ & $5$ & $9$ & $131$\\ \hline
$5$ & $5$ & $5$ & $3$ & $211$\\ \hline
\end{tabular}
\end{table}

\renewcommand{\arraystretch}{1.10}
\begin{table}[htbp]
\centering
\caption{All Dual BiD Codes of Length $729$}
\label{table:dual_bid_729}
\begin{tabular}{|c|c|c|c|c|}
\hline
\hline
$m$ & $r_1$ & $r_2$ & $\dmin$& $K$ \\
\hline
\hline
$6$ & $0$ & $0$ & $2$ & $728$\\ \hline
$6$ & $0$ & $1$ & $4$ & $716$\\ \hline
$6$ & $0$ & $2$ & $8$ & $656$\\ \hline
$6$ & $0$ & $3$ & $16$ & $496$\\ \hline
$6$ & $0$ & $4$ & $32$ & $256$\\ \hline
$6$ & $0$ & $5$ & $64$ & $64$\\ \hline
$6$ & $1$ & $1$ & $3$ & $717$\\ \hline
$6$ & $1$ & $2$ & $8$ & $657$\\ \hline
$6$ & $1$ & $3$ & $16$ & $497$\\ \hline
$6$ & $1$ & $4$ & $32$ & $257$\\ \hline
$6$ & $1$ & $5$ & $64$ & $65$\\ \hline
$6$ & $1$ & $6$ & $729$ & $1$\\ \hline
$6$ & $2$ & $2$ & $6$ & $669$\\ \hline
$6$ & $2$ & $3$ & $16$ & $509$\\ \hline
$6$ & $2$ & $4$ & $32$ & $269$\\ \hline
$6$ & $2$ & $5$ & $64$ & $77$\\ \hline
$6$ & $2$ & $6$ & $243$ & $13$\\ \hline
$6$ & $3$ & $3$ & $7$-$12$ & $569$\\ \hline
$6$ & $3$ & $4$ & $23$-$32$ & $329$\\ \hline
$6$ & $3$ & $5$ & $63$-$64$ & $137$\\ \hline
$6$ & $3$ & $6$ & $81$ & $73$\\ \hline
$6$ & $4$ & $4$ & $8$-$24$ & $489$\\ \hline
$6$ & $4$ & $5$ & $27$ & $297$\\ \hline
$6$ & $4$ & $6$ & $27$ & $233$\\ \hline
$6$ & $5$ & $5$ & $8$-$9$ & $537$\\ \hline
$6$ & $5$ & $6$ & $9$ & $473$\\ \hline
$6$ & $6$ & $6$ & $3$ & $665$ \\ \hline
\end{tabular}
\end{table}
% % % % % % % % % % % % % % % % % 

\subsubsection{Bounds on the Dual Distance of BiD Codes.}

The frequency weight set of a dual BiD code is of the form $\Wc=\{0,\dots,r_1-1\} \cup \{r_2+1,\dots,m\}$. When using Theorem~\ref{thm:dmin_recursion_bounds} on a dual BiD code, the recursion step works with codes of length $3^{m-1}$ with frequency weight sets $\Wc_x \cup \Wc_y$, $\Wc_x \cap \Wc_y$, $\Wc_x$ and $\Wc_y$, all of which are frequency weight sets of dual BiD codes. 
We use the following to terminate the recursion: 
\begin{itemize}
\item[\emph{(i)}] $d_m(\{0,\dots,r\})=3^{m-r}$ (dual Berman code);
\item[\emph{(ii)}] $d_m(\{r,\dots,m\})=2^r$ (Berman code);
\item[\emph{(iii)}] $d_m\left( \{0,\dots,m\} \setminus \{2\}\right)=6$ for $m \geq 4$ (Theorem~\ref{thm:bidd_dmin}); 
\item[\emph{(iv)}] $d_3\left( \{0,1,3\}\right)=5$ (Corollary~\ref{cor:bidd_322_dmin}); and
\item[\emph{(v)}] $d_m\left( \{0,\dots,m\} \setminus \{1\}\right)=3$ for $m \geq 1$;
\end{itemize}
The last result above can be easily proved by showing that no polynomial $f$ of weight $1$ or $2$ satisfies the conditions $\hat{f}_{\p{j}}=0$ for all $\p{j} \in \Zb_3^m$ with $\wt(\p{j})=1$ while the weight-$3$ polynomial $f=1 + X_1\cdots X_m + X_1^2 \cdots X_m^2$ does satisfy these constraints.

The results of this numerical computation are presented in Tables~\ref{table:dual_bid_9_to_243} and~\ref{table:dual_bid_729}. 
The tables show the lower and upper bounds on $\dmin$ of $\bidd(m,r_1,r_2)$ for all codes with $m=2,\dots,6$, and the exact value of $\dmin$ when these two bounds are equal. 

Since $\bidd(m,r_1,r_2)$ is the direct sum of $\abelian(m,\{0,\dots,r_1-1\})$ and $\abelian(m,\{r_2+1,\dots,m\})$, its distance is upper bounded by the minimum of the distances of the latter two codes, i.e., $\min\{3^{m-r_1+1},2^{r_2+1}\}$.
This equals the exact minimum distance for all but $19$ codes in Tables~\ref{table:dual_bid_9_to_243} and~\ref{table:dual_bid_729}.

\subsection{Fast Decoding of $\bid(m,1,1)$ and $\bid(m,0,1)$} \label{app:decoding_first_order}

The generator matrices of BiD codes enjoy a recursive structure~\cite[Lemma~2.3]{DNNK_ITW_25}, that expresses the code of length $3^m$ in terms of codes of length $3^{m-1}$; this is similar to the Plotkin construction of RM codes. 
In the case of $\bid(m,1,1)$ and $\bid(m,0,1)$, this recursion resembles the structure of a family of codes known as first-order \emph{recursive subproduct codes}~\cite{SNK_ISIT24}. 
Under appropriate modifications, the fast implementations of the ML and the max-log-MAP decoders of first-order recursive subproduct codes from~\cite[Section~III]{SNK_ISIT24} apply to $\bid(m,1,1)$ and $\bid(m,0,1)$. 
We now explain how the latter two codes fit into the decoding framework of~\cite[Section~III]{SNK_ISIT24}.

\subsubsection{Decoding $\bid(m,1,1)$.} \label{sec:sub:decoding_bid_m11}

Let $\p{G}_{m,w}$ denote the generator matrix of $\bid(m,w,w)$. We use the following two facts from Lemma~2.3 of~\cite{DNNK_ITW_25}: 
\emph{(i)}~$\p{G}_{m,1}$ is the vertical stacking of the following three matrices, $(1,1,1) \otimes \p{G}_{m-1,1}$, $(1,1,0) \otimes \p{G}_{m-1,0}$ and $(1,0,1) \otimes \p{G}_{m-1,0}$; and
\emph{(ii)}~$\bid(m-1,0,0)$ is the repetition code, i.e., $\p{G}_{m-1,0}$ is the $1 \times 3^{m-1}$ matrix consisting of all ones.
Thus, for any $\p{c} \in \bid(m,1,1)$ there exist $\p{d} \in \bid(m-1,1,1)$ and $b_1,b_2 \in \Fb_2$ such that 
\begin{equation*}
\p{c} = (1,1,1) \otimes \p{d} + b_1 (1,1,0) \otimes \p{1}_{3^{m-1}} + b_2 (1,0,1) \otimes \p{1}_{3^{m-1}},
\end{equation*}
where $\p{1}_{\ell}$ denotes the all-ones vector of length $\ell$. 
This map from $\bid(m,1,1)$ to $\bid(m-1,1,1) \times \Fb_2 \times \Fb_2$ is a one-to-one correspondence.
Denoting $(b_1+b_2,b_1,b_2)$ by $\p{a}$, we have
\begin{equation*}
\p{c} = \p{1}_3 \otimes \p{d} + \p{a} \otimes \p{1}_{3^{m-1}}.
\end{equation*}
We observe that, up to a permutation of coordinates, the above expression is identical to~\cite[equation~(4)]{SNK_ISIT24} (with the code $\bid(m-1,1,1)$ playing the role of $\mathscr{C}^{\otimes[1,m-1]}$ from~\cite{SNK_ISIT24}, and the integer $3$ taking the place of $n$ in~\cite{SNK_ISIT24}).
With this correspondence established, we can directly apply the efficient ML and max-log-MAP decoders from~\cite{SNK_ISIT24} to $\bid(m,1,1)$.
The main idea behind these decoders is to decode the length-$3^{m-1}$ codeword $\p{d}$ for each of the four possibilities of $\p{a}$, and then pick the best choice of $(\p{d},\p{a})$ from among these four pairs.

\subsubsection{Decoding $\bid(m,0,1)$.}

Applying the direct-sum decomposition property to $\abelian(m,\{0,1\})= \bid(m,0,1)$ we observe that $\bid(m,0,1) = \bid(m,1,1) \oplus \bid(m,0,0)$. 
Following the same argument as in Section~\ref{sec:sub:decoding_bid_m11} along with the fact that $\bid(m,0,0)$ is the repetition code, we see that every $\p{c'} \in \bid(m,0,1)$ can be expressed as 
\begin{equation*}
\p{c'} = \p{1}_3 \otimes \p{d} + \p{a} \otimes \p{1}_{3^{m-1}} + b_3 \p{1}_{3^m},
\end{equation*}
where $b_3 \in \Fb_2$ and $\p{d},\p{a}$ are as in Section~\ref{sec:sub:decoding_bid_m11}. Since $\p{1}_{3^m} = \p{1}_3 \otimes \p{1}_{3^{m-1}}$ and $\bid(m-1,0,1)$ is the direct sum of $\bid(m-1,1,1)$ and the repetition code of length $3^{m-1}$, 
we have 
\begin{equation*}
\p{1}_3 \otimes \p{d} + b_3 \p{1}_{3^m} = \p{1}_3 \otimes \left( \p{d} + b_3 \p{1}_{3^{m-1}} \right) = \p{1}_3 \otimes \p{d'}    
\end{equation*}
where $\p{d'} \in \bid(m-1,0,1)$.
Hence, $\p{c'} = \p{1}_3 \otimes \p{d'} + \p{a} \otimes \p{1}_{3^{m-1}}$, which puts us into the recursive decoding framework of~\cite{SNK_ISIT24}.

\subsection{Belief Propagation Decoding of $\bid(m,2,2)$} \label{app:bp_decoding}

We described the factor graph in Section~\ref{sec:sub:BP_decoding}.
We use the standard check node update rule at all degree-$6$ check nodes (without the min-sum approximation).
As in~\cite{LHP_ISIT_20}, we use weighted message-passing decoding where the messages arriving at the variable nodes originating from the degree-$6$ check nodes, generalized check nodes with $\ell=1$, and those with $\ell=2$, are scaled by carefully chosen weights $\beta_0$, $\beta_1$ and $\beta_2$, respectively.
While the values of these weights can be optimized for each value of $m$ as well as for the channel conditions, we use the constants $\beta_0=0.075$, $\beta_1=0.0375$, $\beta_2=0.0075$ in all our simulations.

We use an interleaved schedule of passing messages, where we operate all the generalized check nodes with \mbox{$\ell=1$} and those with \mbox{$\ell=2$} alternately, and operate all the degree-$6$ check nodes between the schedules for the $\ell=1$ and $\ell=2$ nodes. 
For each of these three classes of check nodes (degree-$6$ check nodes, generalized check nodes with $\ell=1$, or generalized check nodes with $\ell=2$), operations on all nodes in the class are performed in parallel, and these operations are immediately followed by updates at all the variable nodes, so that the updated information is available to the next class of check nodes in the interleaved schedule.
% Further, we update all the variable nodes immediately after receiving updates from any of the check nodes or generalized check nodes.

\subsubsection*{Complexity Analysis.}
% A straightforward analysis shows that the per-iteration complexity order of this decoder is $N^{\log6 / \log 3} \log N=N^{1.63} \log N$, which is the order of the number of degree-$6$ check nodes.
The number of generalized check nodes corresponding to projections with \mbox{$\ell=1$} is $3m$ and the complexity of soft-in soft-out decoding at each of these generalized check nodes is $\Theta(N^{1.26})$. Similarly, the number of generalized check nodes with $\ell=2$ is $18 \binom{m}{2}$ and the complexity order of processing each of them is $N^{1.26}$. Since $m=\log_3 N$, the overall complexity order of operating all the generalized check nodes is $N^{1.26} \log^2 N$. 
Now, the number of degree-$6$ check nodes is $m2^{m-2}3^{m-1}=m6^m/12$, which is of the order $N^{1.63} \log N$. Since the complexity of processing each of these check nodes is independent of $N$, the complexity of processing all the degree-$6$ check nodes is $\Theta(N^{1.63} \log N)$. 
Since $\log N$ grows slower than $N^{\delta}$ for any $\delta>0$, we see that the overall complexity order of the BP decoder per iteration is $N^{1.63} \log N$. 

\subsubsection*{Code corresponding to the factor graph.}

The linear code defined by a factor graph is the set of binary vectors that satisfy all the constraints imposed by the check nodes and the generalized check nodes present in the graph. 
In the case our BP decoder, the degree-$6$ check nodes constrain this linear code to be $\bid(m,2,2) \oplus \bid(m,0,0)$ (see Appendix~\ref{app:span_of_min_wt_codewords}). 
The generalized check nodes do not impose any further restrictions: using the fact that $\bid(m,0,0)$ is the repetition code and using Theorem~\ref{thm:projection} it is easy to show that for any $f \in \bid(m,2,2) \oplus \bid(m,0,0)$ we have 
\begin{equation*}
\punc(\p{f},S,\p{u}) + \punc(\p{f},S,\p{v}) \in \bid(m-\ell,w-\ell,w-1),
\end{equation*}
where $S \subset \{1,\dots,m\}$, $|S|=\ell \leq w$ and $\p{u},\p{v} \in \Zb_3^{\ell}$.
Thus, the code corresponding to the factor graph used by our BP decoder is the super-code $\bid(m,2,2) \oplus \bid(m,0,0) \supsetneq\bid(m,2,2)$. The dimension of this super-code is one more than the dimension of $\bid(m,2,2)$.

\begin{figure}[t!]
    \centering
    \includegraphics[width=\linewidth]{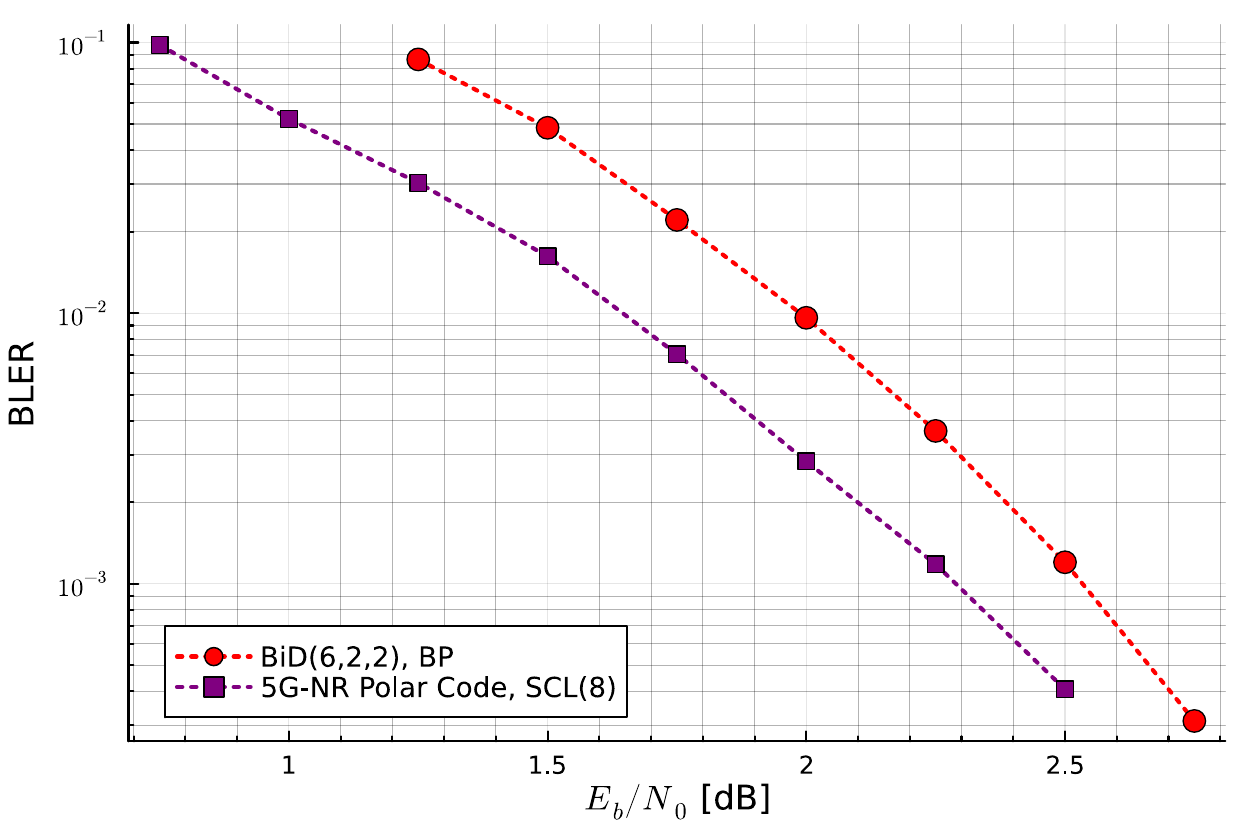}
    \caption{BLER comparison of codes with length $729$}
    \label{fig:729}
\end{figure}

\subsection{BLER Comparison of Codes of Length $729$} \label{app:fig:729}

Fig.~\ref{fig:729} compares two codes of length $729$ with the same rate, viz., $\bid(6,2,2)$ under BP decoding and the 5G-NR uplink CRC-aided Polar code under SCL($8$)~\cite{Matlab_Polar}. The 5G-NR code uses a $11$-bit CRC for data protection and length adaptation to achieve a block length of $729$. 
The BiD code is about $0.25$~dB worse at BLER approximately $10^{-3}$. 
We were unable to simulate the ML decoding performance of these codes because of the high complexity.

% % % % % % % % % % 
\hfill\IEEEQED

\end{document}